\documentclass[sn-apa]{sn-jnl}  

\usepackage{graphicx}
\usepackage{multirow}
\usepackage{amsmath,amssymb,amsfonts}
\usepackage{amsthm}
\usepackage{mathrsfs}
\usepackage[title]{appendix}
\usepackage{xcolor}
\usepackage{textcomp}
\usepackage{manyfoot}
\usepackage{booktabs}
\usepackage{algorithm}
\usepackage{algorithmicx}
\usepackage{algpseudocode}
\usepackage{listings}
\usepackage{soul}
\usepackage{subcaption}

\begin{document}

\title[The Evolution of the Peridynamics Community]{The Evolution of the Peridynamics Community in Its First Quarter Century}

\author[1]{\fnm{Biraj} \sur{Dahal}}\email{bdahal6@gatech.edu}

\author*[2]{\fnm{Pablo} \sur{Seleson}}\email{selesonpd@ornl.gov}

\affil[1]{\orgdiv{School of Mathematics}, \orgname{Georgia Institute of Technology}, \orgaddress{\street{North Ave NW}, \city{Atlanta}, \state{GA}, \postcode{30332}, \country{USA}}}

\affil[2]{\orgdiv{Computer Science and Mathematics Division}, \orgname{Oak Ridge National Laboratory}, \orgaddress{\street{P.O. Box 2008, MS-6013}, \city{Oak Ridge},  \state{TN}, \postcode{37831-6013}, \country{USA}}}

\abstract{Peridynamics is a fast growing field of continuum mechanics, especially developed for the modeling and simulation of fracture problems, initiated a quarter century ago. In this study, we analyze the evolution of the peridynamics community since its inception in terms of publication co-authorship. For this purpose, we construct a peridynamics co-authorship network for each year from 2000 to 2024 and perform network analysis based on selected metrics. Nodes represent scientists, and links connect co-authoring scientists with link weights representing the number of co-authorships (based on the total number of co-authors per publication). Network-level metrics are used to quantify the evolution of the field, and node-level metrics are used to identify trends in the most collaborative scientists in peridynamics. We noticed a deviation in network trends that occurred in the years since 2019, and we subsequently performed a  country-based analysis with insights about the impact of the COVID-19 pandemic on the evolution of the peridynamics co-authorship network.
}


\maketitle

\section{Introduction}
Peridynamics is a field of continuum mechanics developed in 2000, which uses a non-local formulation that is especially suitable for modeling fracture~\citep{silling2000reformulation}. Over the past quarter century, the field has grown to encompass thousands of scientists and research publications worldwide. In order to quantify the evolution of the peridynamics community, we employ network analysis to study the peridynamics co-authorship network from 2000 to 2024. 
We construct a network for each year whose nodes represent authors of peridynamics publications, and whose links connect two nodes if the corresponding authors co-authored a peridynamics publication by that given year. Using this approach, we analyze the evolving network over the years. 
The peridynamics co-authorship network grew from 1 node in the year 2000 to thousands of nodes and links in the year 2024. Network-level metrics are used to glean insights about the community as a whole (see Section \ref{sec:networklevel}), and node-level metrics are used to identify the most collaborative scientists (see Section \ref{sec:nodelevel}).

In this paper, we analyze the peridynamics co-authorship network with an emphasis on 2020-2024, an extension of our previous analysis of the peridynamics co-authorship network until 2019~\citep{dahal2023evolution}. The major differences between the current analysis and the previous analysis are discussed in Section \ref{sec:difference}. In the years since 2019, we noticed a deviation in network trends observed in our previous work, especially when considering the countries of the authors' institutional affiliations, which seems to be a result of the COVID-19 pandemic; see Section \ref{sec:country} for more details.

The rest of this paper is organized as follows. In the remainder of this section, we explain the data collection process (Section~\ref{sec:data}) and differences from our previous work (Section~\ref{sec:difference}). In Section~\ref{sec:networklevel}, we perform a network-level analysis of the peridynamics co-authorship network. In Section~\ref{sec:nodelevel}, we perform a node-level analysis of the peridynamics co-authorship network. In Section~\ref{sec:country}, we analyze the countries of the institutional affiliations of new scientists joining the peridynamics co-authorship network. Finally, in Section~\ref{sec:summary}, we summarize the main implications of this study.

\subsection{Data collection}
\label{sec:data}
We obtained our dataset (henceforth called the publication-author list) as the collection of all the  publications indexed by Scopus ~\citep{scopus} published no later than 2024 that have ``{\it peridynamic}'' or ``{\it peridynamics}'' in the title, abstract, or author keywords. The data was downloaded on August 13, 2025. Since the Scopus databases are continuously updated with new entries, the publication-author list might be missing a few peridynamics publications that were published by 2024 but were not indexed by Scopus at the time of data download. Nevertheless, we expect the impact of such a potential omission on the network analysis results to be negligible, especially for years prior to 2024. We also capped the number of authors for each publication at 10 (taking only the first 10 listed authors when applicable), which affects six publications.

From the publication-author list, we constructed co-authorship networks representing each year. For every year $y$ from 2000 to 2024, the network $G_y$ is constructed using only data from the publication-author list published on or before the year $y$. $G_y$'s nodes represent peridynamics authors, and $G_y$'s links connect two nodes if the corresponding authors co-authored a publication in the publication-author list. The links are weighted according to the fractional counting scheme~\citep{perianes2016constructing}, meaning that each publication with $k$ authors contributes $\frac{1}{k-1}$ to the weight of each of the ${k \choose 2}$ links connecting the nodes representing those authors.

We remark that we used the unique author IDs provided by Scopus to construct the co-authorship network. We found that the provided author IDs were sufficient for author-name disambiguation, except for one mistake. The author identified as ``Askari, E." and ``Askari, A." was given two separate Scopus IDs. We had to manually combine their records as a pre-processing step. This was the only modification we applied to the author IDs provided by Scopus.

\subsection{Differences from the previous analysis}
\label{sec:difference}
This paper applies network analysis to study the peridynamics co-authorship network up to 2024, similar to the analysis in \cite{dahal2023evolution} that studied years up to 2019. We employ a subset of the previous analysis, which we believe is most relevant to the peridynamics community as a whole. A more detailed and granular network analysis can be found in \cite{dahal2023evolution}. We also incorporate a new type of analysis based on the countries of the institutional affiliations of the authors listed in the publication-author list.

In addition, we use Scopus instead of Web of Science \citep{wos}, which was used in our previous work. We chose Scopus because it has a larger dataset of peridynamics publications: we obtained records for 2761 different documents from Scopus, while there were only 2197 different documents found in Web of Science. There were 614 publications in the former database that were not in the latter, while only 50 publications were found in Web of Science but not in Scopus (meaning they have 2147 publications in common). In addition, Scopus provides unique author IDs, which are an automatic form of author-name disambiguation. In contrast, Web of Science does not provide author IDs, so one would have to employ an author-name disambiguation procedure (which may require a significant amount of manual disambiguation), as was done in our previous work.

\section{Network-level Analysis}\label{sec:networklevel}
In this section, we examine network-level metrics for the peridynamics co-authorship network. For a more comprehensive discussion on network-level metrics, see \cite{dahal2023evolution}.

\subsection{Data size}\label{sec:datasize}
\begin{figure}[h!]
    \centering
    \includegraphics[width=0.5\linewidth]{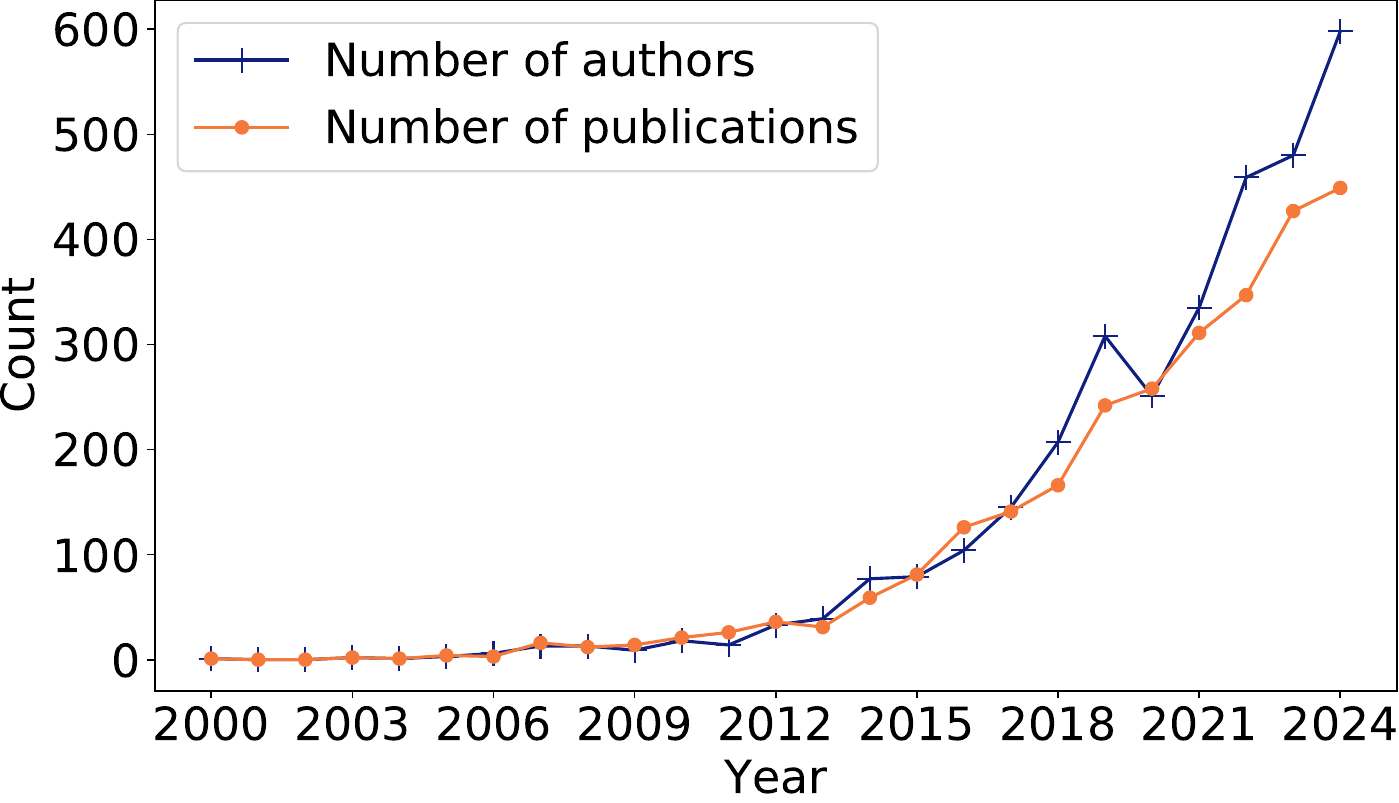}
    \caption{Number of new authors and publications per year.}
    \label{fig:size}
\end{figure}

\begin{figure}[h!]
    \centering
     \begin{subfigure}[b]{0.5\textwidth}
     \centering
    \includegraphics[width=\linewidth]{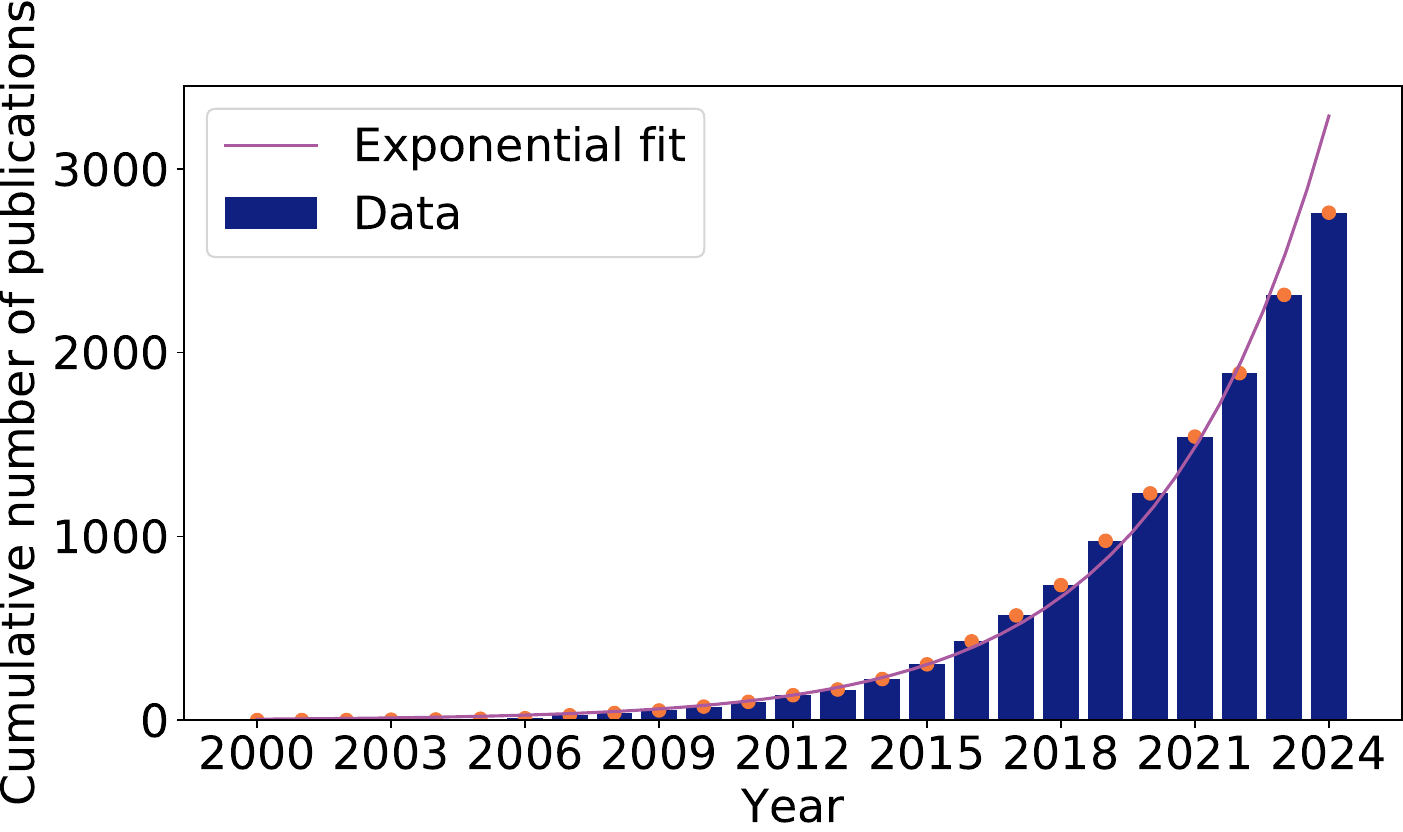}
    \caption{Number of publications}
    \label{fig:numpapers}
    \end{subfigure}%
    ~
    \begin{subfigure}[b]{0.5\textwidth}
     \centering
    \includegraphics[width=0.98\linewidth]{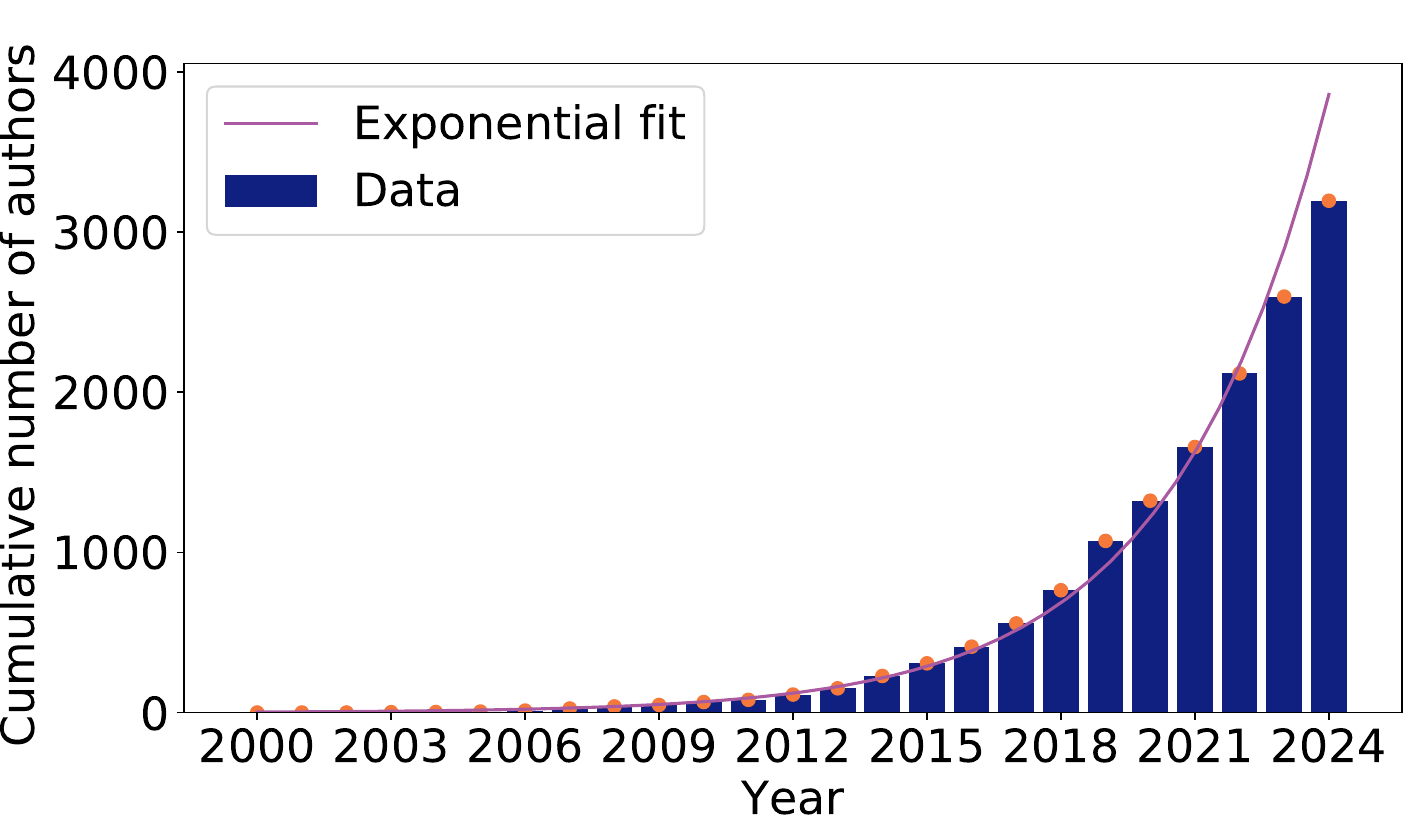}
    \caption{Number of authors}
    \label{fig:numauthors}
    \end{subfigure}%
    \caption{Cumulative number of publications and authors per year.}
    \label{fig:numpapersauthors}
\end{figure}

Data size refers to metrics only involving the number of authors and publications. Figure~\ref{fig:size} shows the number of new authors and publications in the publication-author list per year. By 2024, there were a total of 2761 peridynamics publications and 3195 peridynamics authors. Figures~\ref{fig:numpapers} and~\ref{fig:numauthors} show the cumulative number of peridynamics publications and authors, respectively, per year with exponential fits (the fits used data from 2010-2024). We see that the growth in the number of publications and authors is overestimated by the exponential fits in 2023 and 2024.

\begin{figure}[h!]
    \centering
    \includegraphics[width=0.5\linewidth]{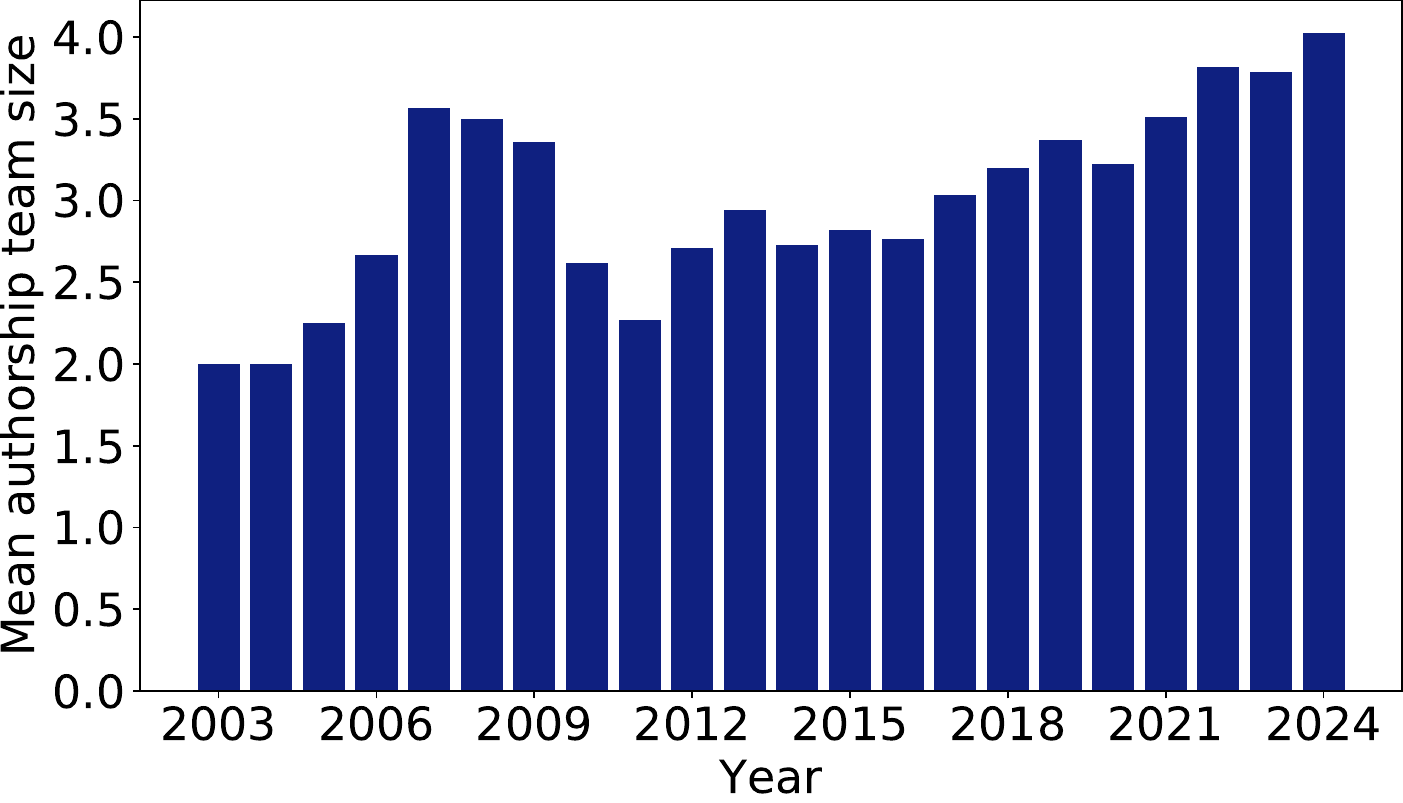}
    \caption{Mean authorship team size per year.}
    \label{fig:teamsize}
\end{figure}

\begin{figure}[h!]
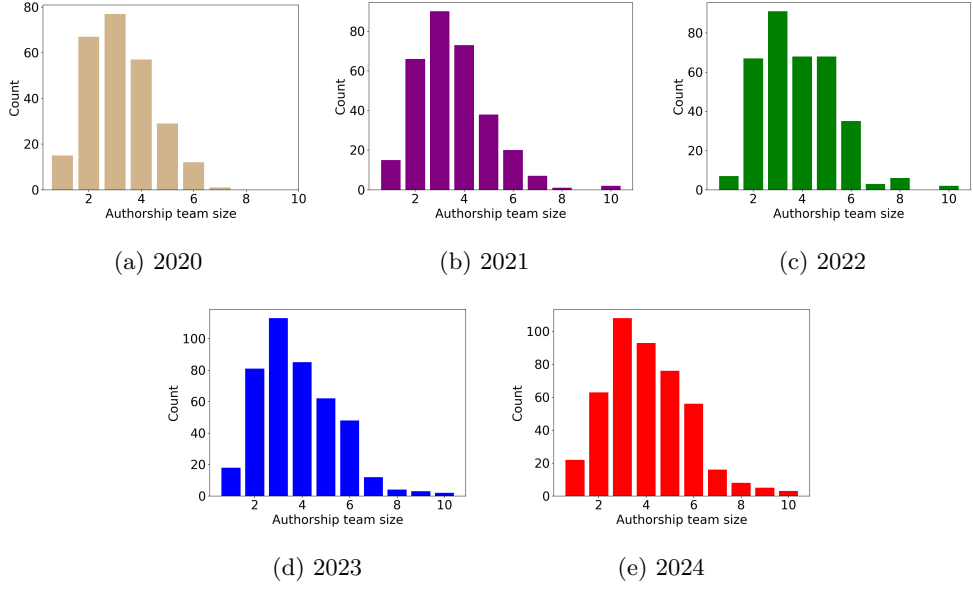

\centering
     \begin{subfigure}[b]{0.33\textwidth}
     \centering
    \includegraphics[width=0.99\linewidth]{figures/teamsize-2020.pdf}
    \caption{2020}
    \end{subfigure}%
    \centering
     \begin{subfigure}[b]{0.33\textwidth}
     \centering
    \includegraphics[width=0.99\linewidth]{figures/teamsize-2021.pdf}
    \caption{2021}
    \end{subfigure}%
    ~
    \begin{subfigure}[b]{0.33\textwidth}
     \centering
    \includegraphics[width=0.99\linewidth]{figures/teamsize-2022.pdf}
    \caption{2022}
    \end{subfigure}
    \\
    \begin{subfigure}[b]{0.33\textwidth}
    \centering
    \includegraphics[width=0.99\linewidth]{figures/teamsize-2023.pdf}
    \caption{2023}
    \end{subfigure}%
    ~
    \begin{subfigure}[b]{0.33\textwidth}
     \centering
    \includegraphics[width=0.99\linewidth]{figures/teamsize-2024.pdf}
    \caption{2024}
    \end{subfigure}%
    \caption{Distribution of authorship team sizes from 2020 to 2024.}
    \label{fig:teamsizedistribution}
\end{figure}

Figure \ref{fig:teamsize} shows the mean authorship team size, or average number of authors per publication, for each year; note that this plot is not cumulative. Accounting for all publications from 2000 to 2024, the mean authorship team size is 3.48. We plot the frequency of authorship team sizes for each year from 2020 to 2024 in Figure \ref{fig:teamsizedistribution}. The overall distribution of authorship team sizes is roughly similar for those years.

\subsubsection{Data size discussion}
The peridynamics community has grown rapidly, almost exponentially, since its inception in 2000, and this trend has continued all the way to 2024 (see Figures~\ref{fig:size} and \ref{fig:numpapersauthors}). However, an exponential fit overestimates the growth in 2023 and 2024 (see Figure~\ref{fig:numpapersauthors}), which indicates that there may have been a slowdown in recent years. In particular, we notice a significant deviation in the growth of the number of new authors with a drop in 2020, together with a slowdown in the number of new publications, compared to 2019 (see Figure~\ref{fig:size}). We examine this potential slowdown in more detail in Section~\ref{sec:country}. Nevertheless, by 2024 the number of peridynamics publications has more than doubled (increasing by a factor of 2.8) and the number of peridynamics authors has tripled, compared to 2019, when there were 990 peridynamics publications and 1072 authors.

The average authorship team size has been on an overall increasing trend since at least 2014, with the 2024 mean authorship team size being 4.02 (see Figure~\ref{fig:teamsize}). From 2020 to 2024, the most common authorship team size was three, but about half (49.9\%) of author teams across those years combined are larger than that (see Figure~\ref{fig:teamsizedistribution}). In addition, many publications with large author teams (e.g., five or more authors) have increasingly been published in recent years, compared to previous years (see Figure~\ref{fig:teamsizedistribution}). For example, only 127 out of 990 publications (or 12.8\%) published on or before 2019 had five or more authors. In contrast, 518 out of 1771 publications (or 29.2\%) published during or after 2020 had five or more authors, which is more than double.

\subsection{Connected components}
The peridynamics co-authorship network is a single connected component from 2000 to 2005, but there are three connected components in 2006, two connected components in 2007, four connected components in 2008 and 2009, and the number of connected components increases each year thereafter. A connected component of a network is a maximally connected subnetwork: all nodes within a connected component are linked directly or via a path of links, but there are no links between nodes in two different connected components.

\begin{figure}[h!]
    \centering
    \fbox{\includegraphics[width=0.5\linewidth]{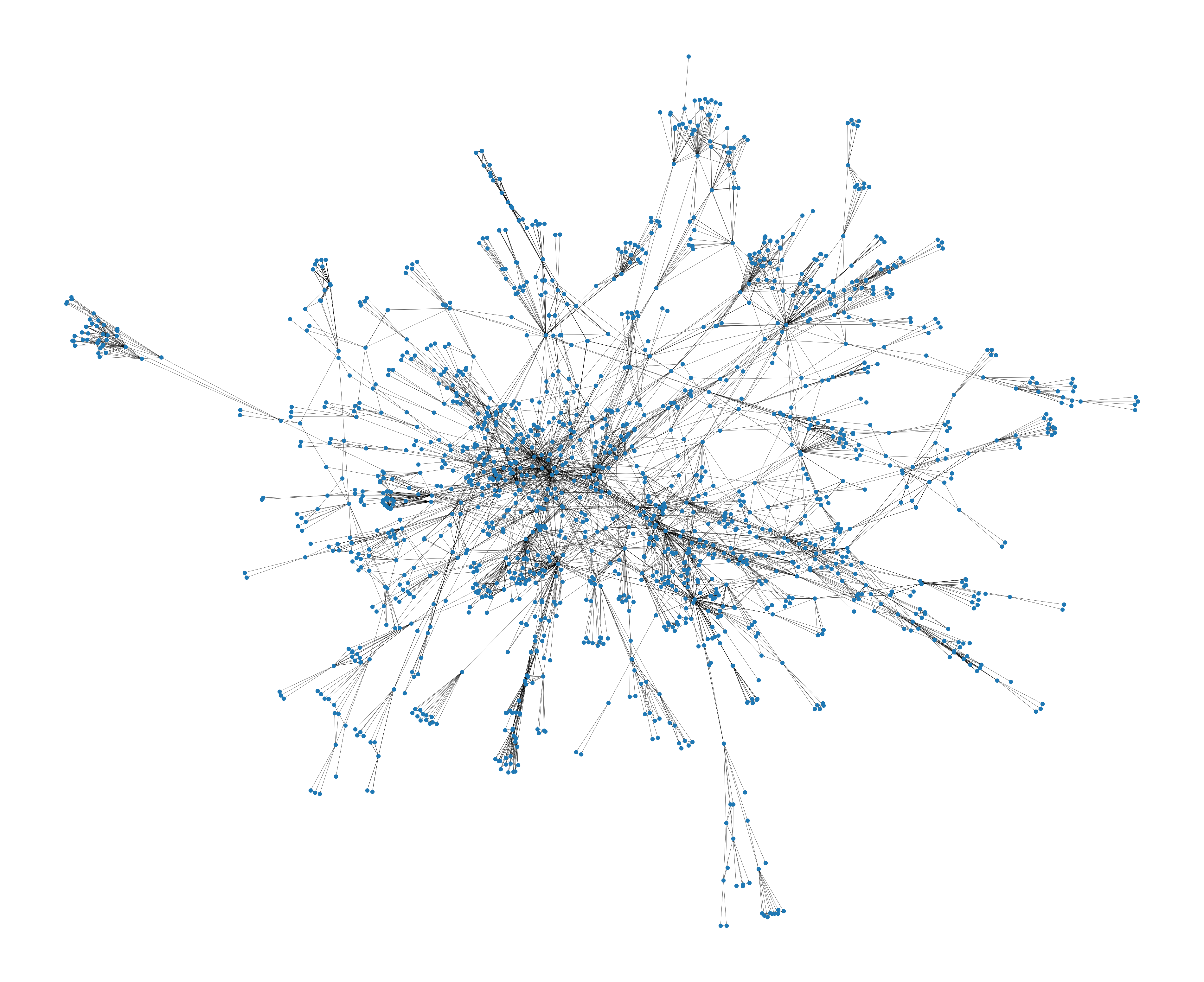}}
    \caption{Visualization of the largest connected
component (LCC) for 2024.}
    \label{fig:lcc}
\end{figure}

\begin{figure}[h!]
    \centering
    \includegraphics[width=0.5\linewidth]{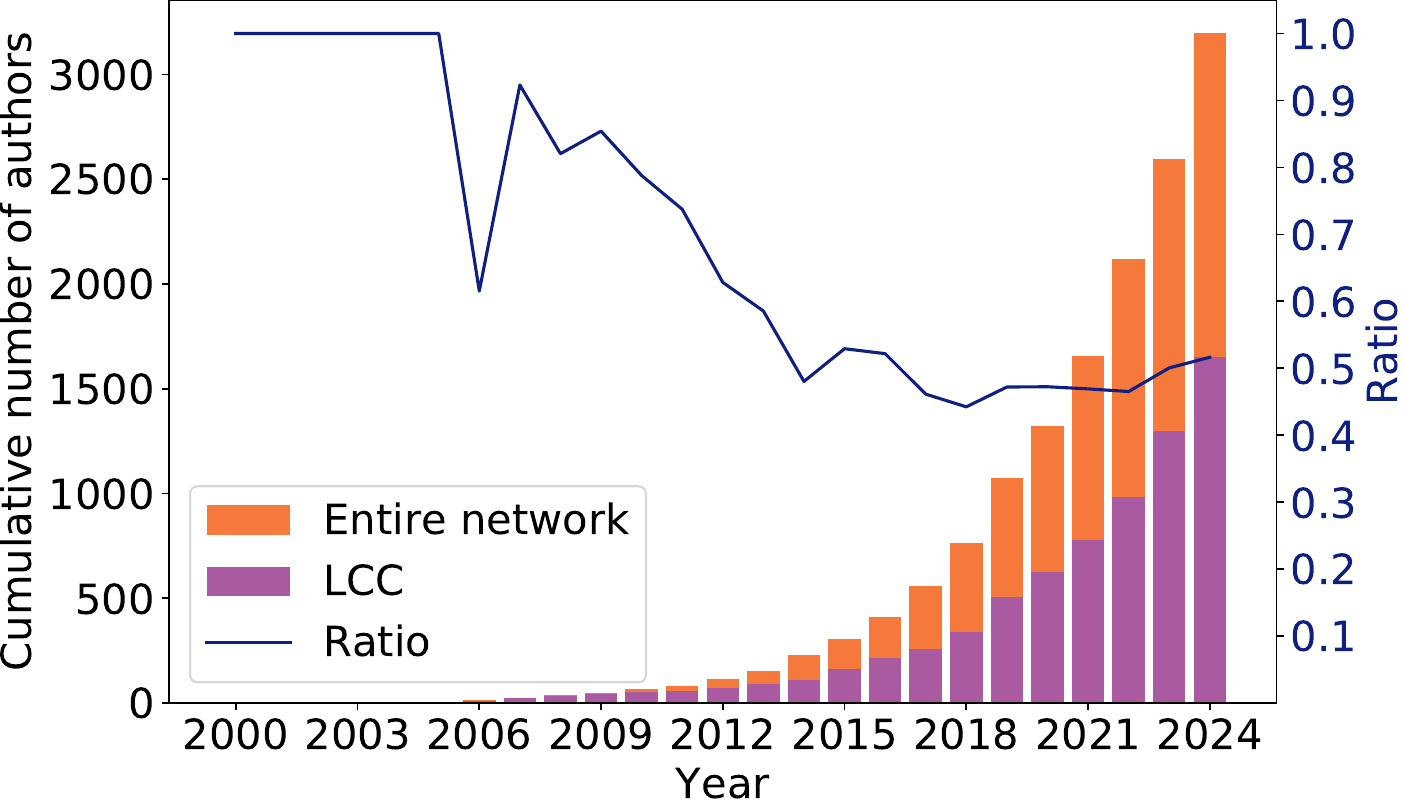}
    \caption{Size of the largest connected
component (LCC) and the entire network per year.}
    \label{fig:ratiolcc}
\end{figure}

The peridynamics co-authorship network has a largest connected component (LCC) with 1650 nodes in 2024. We visualize the LCC for 2024 in Figure \ref{fig:lcc}, where the layout is determined by applying the \texttt{spring\_layout} function in NetworkX~\citep{hagberg2008exploring} (which implements the Fruchterman-Reingold algorithm \citep{fruchterman1991graph}). Each blue dot visualizes a node (author), and each gray line segment represents a link (co-authorship) between two nodes. A comparison of the size of the LCC with the size of the entire network for each year is shown in Figure~\ref{fig:ratiolcc}. The LCC comprises about half of the nodes of the entire network in recent years.

A more thorough and interactive visualization of the peridynamics co-authorship network can be obtained using PDnetwork~\citep{dahal_2025_16897786} at: \url{https://ornl.github.io/PDnetwork/}. This web tool allows users to compute some of the metrics described in this paper, and it provides links and searching features for the publication-author list. Note that PDnetwork may use more up-to-date data than was available at the time of writing this paper, so slight differences may exist between the network presented on PDnetwork and the network analyzed in this paper.

\begin{figure}[h!]
    \centering
    \includegraphics[width=0.5\linewidth]{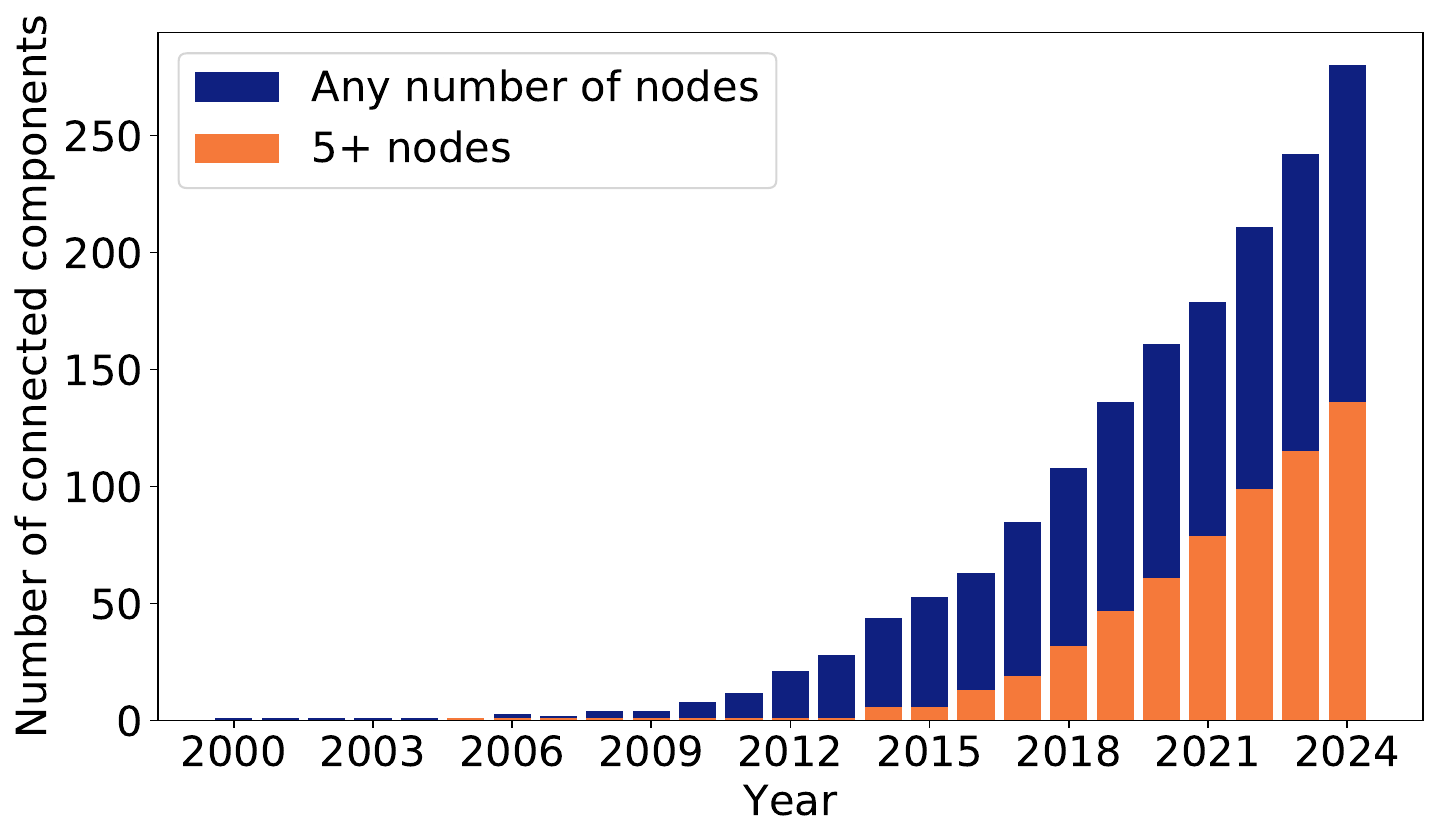}
    \caption{Number of connected components per year.}
    \label{fig:numcomponents}
\end{figure}

The number of connected components in the peridynamics co-authorship network in 2024 is 280, though only 136 of these connected components contain five or more nodes. The number of connected components per year is shown in Figure \ref{fig:numcomponents}, with the number of connected components having at least five nodes indicated in orange. We observe that in all years, at least half of the connected components have fewer than five nodes (except for 2005 when there is only one connected component and it has seven nodes).

\begin{figure}[h!]
    \centering
     \begin{subfigure}[b]{0.45\textwidth}
     \centering
    \includegraphics[width=\linewidth]{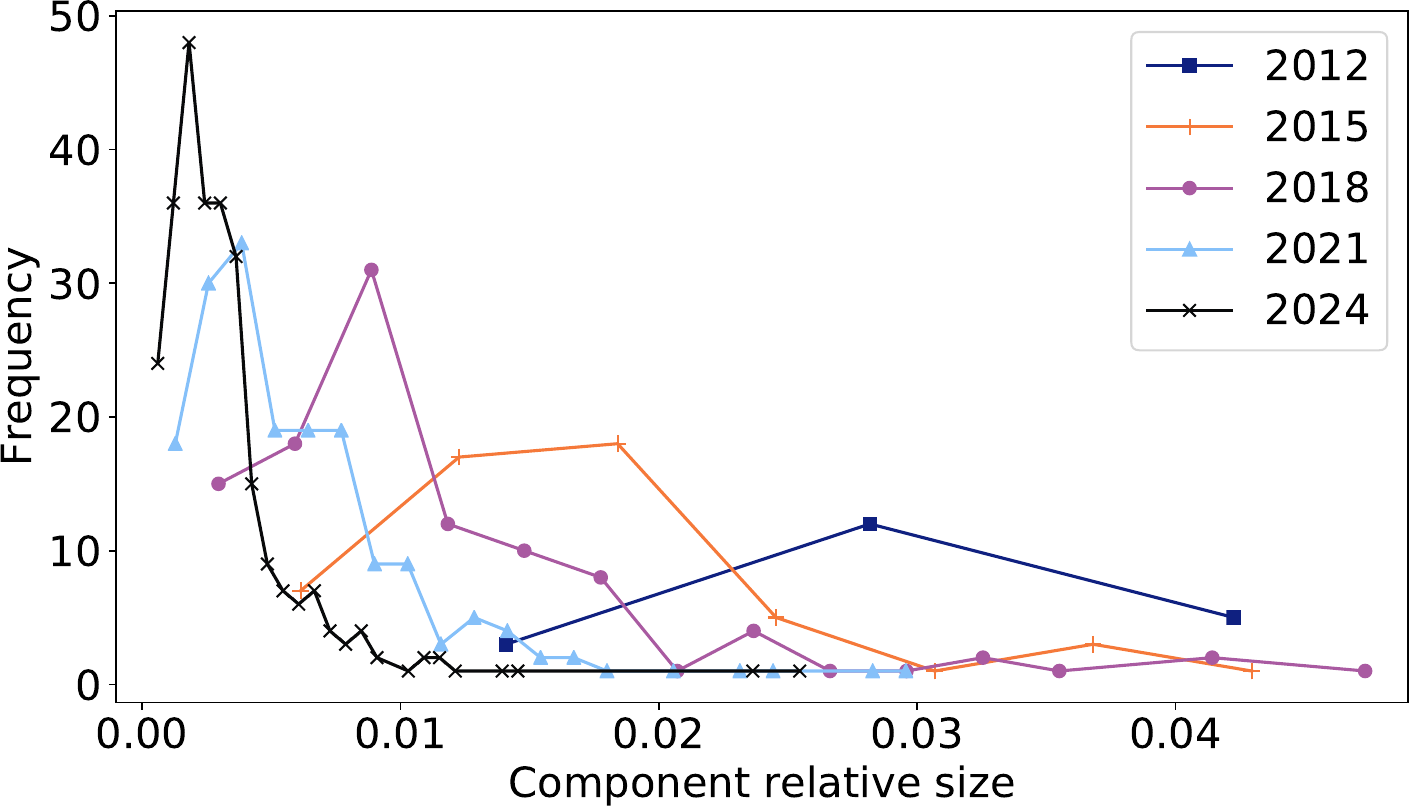}
    \caption{Every three years since 2012}
    \label{fig:componentdisthistorical}
    \end{subfigure}
~
     \begin{subfigure}[b]{0.45\textwidth}
     \centering
    \includegraphics[width=\linewidth]{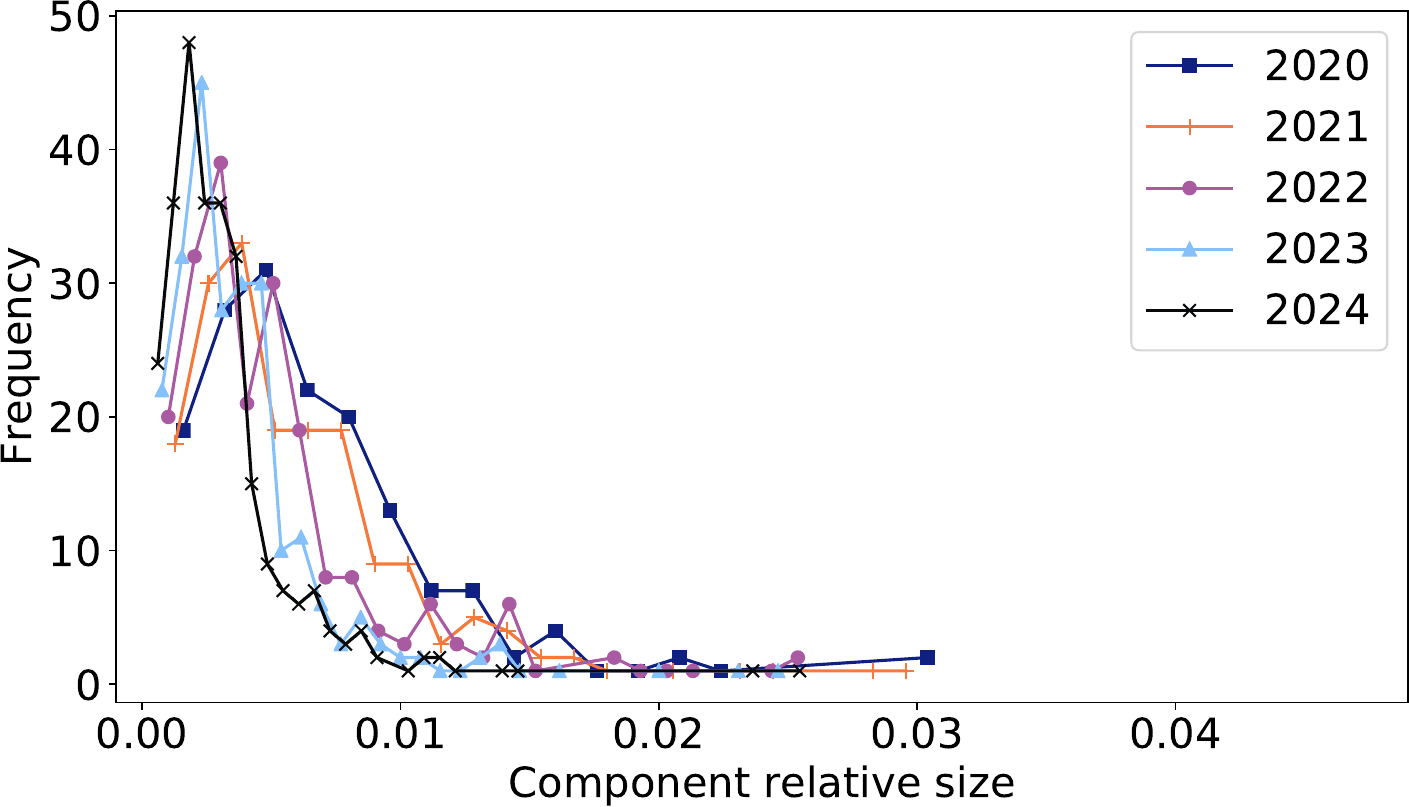}
    \caption{Each year since 2020}
    \label{fig:componentdist}
    \end{subfigure}
    \caption{Relative size distribution of non-largest connected components.}
    \label{fig:componentdist-both}
\end{figure}

The connected components other than the LCC are quite small, with the second-largest connected component consisting of only 42 nodes in 2024 (compared to 1650 nodes in the LCC in 2024). Figure \ref{fig:componentdisthistorical} shows the distribution of the connected components' relative sizes (relative to the LCC size) for every third year since 2012. Similarly, Figure \ref{fig:componentdist} shows the distribution of the connected components' relative sizes for each year since 2020.

\begin{figure}[h!]
    \centering    \includegraphics[width=0.5\linewidth]{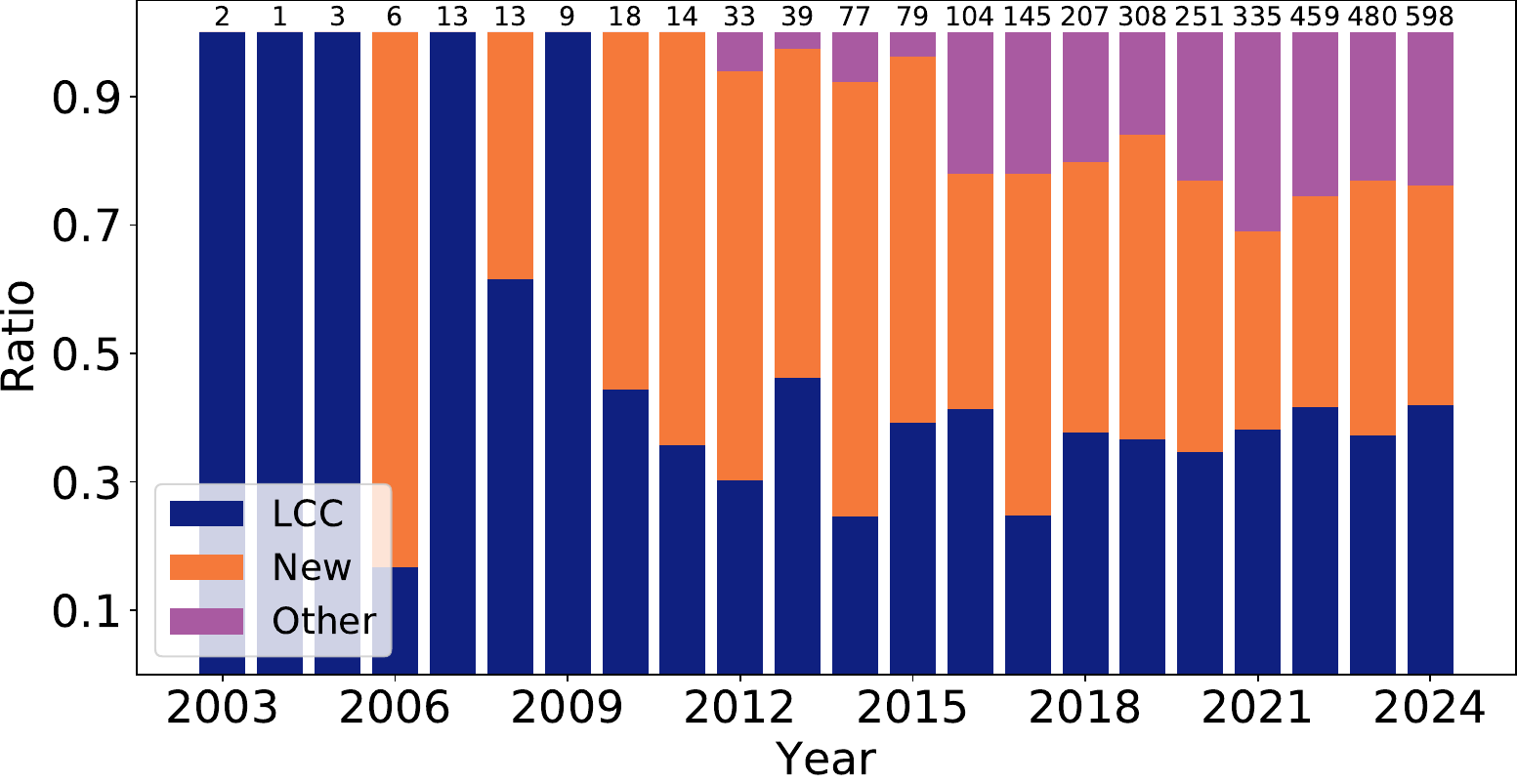}
    \caption{Connected component affiliation of new nodes per year. ``New'' refers to nodes that form a new connected component when they join the network, and ``Other'' refers to nodes that join an existing connected component other than the largest connected component (LCC). The numbers above each bar indicate the number of new nodes that year.}
    \label{fig:newnodes}
\end{figure}

Every year, new nodes join the peridynamics co-authorship network by either linking to a node in the LCC, linking to a node in another connected component, or creating a new connected component. Figure \ref{fig:newnodes} is a stacked bar plot showing the connected component affiliation of new nodes per year. The percentage of nodes joining the LCC increased from 34.7\% in 2020 to 41.6\% in 2022, decreased to 37.3\% in 2023, and then increased to 42.0\% in 2024.

\subsubsection{Connected components discussion}
The LCC represents the core of the peridynamics community (see, e.g.,  Figure~\ref{fig:lcc} for its 2024 visualization) and comprises about 47\% to 52\% of the entire network in each year after 2019 (see Figure~\ref{fig:ratiolcc}), while accounting for at least 44.3\% of the entire network for all years. From 2020 to 2022, the relative size of the LCC slightly decreased, but then increased to 50.2\% in 2024 (see Figure~\ref{fig:ratiolcc}). While the LCC is roughly half the size of the entire network in recent years (e.g., since 2014), the other connected components are much smaller, with only a minority of the connected components having five or more nodes (see Figure~\ref{fig:numcomponents}).

In recent years, we see that the connected component size distribution curves become more right-skewed (see Figure~\ref{fig:componentdist-both}), meaning the curves concentrate on lower values. We also observe that since 2019, the size of the LCC is growing in comparison to the size of the second-largest connected component(s), i.e., the rightmost point for each curve is shifting to the left (see Figure~\ref{fig:componentdist-both}). In other words, as time goes on, the LCC continues to grow in size relative to every other connected component. Note that in 2011, 2012, 2016, 2020, and 2022, there are multiple connected components of equal size that are the second-largest connected components. On the other hand, a sizable portion of scientists join connected components other than the LCC in every year since 2016 (see Figure~\ref{fig:newnodes}). This potentially indicates that there are a lot of sub-communities in peridynamics that are not linked to the LCC, which could be an interesting topic for future exploration.

\subsection{Other metrics}
\begin{figure}[h!]
    \centering
    \includegraphics[width=0.5\linewidth]{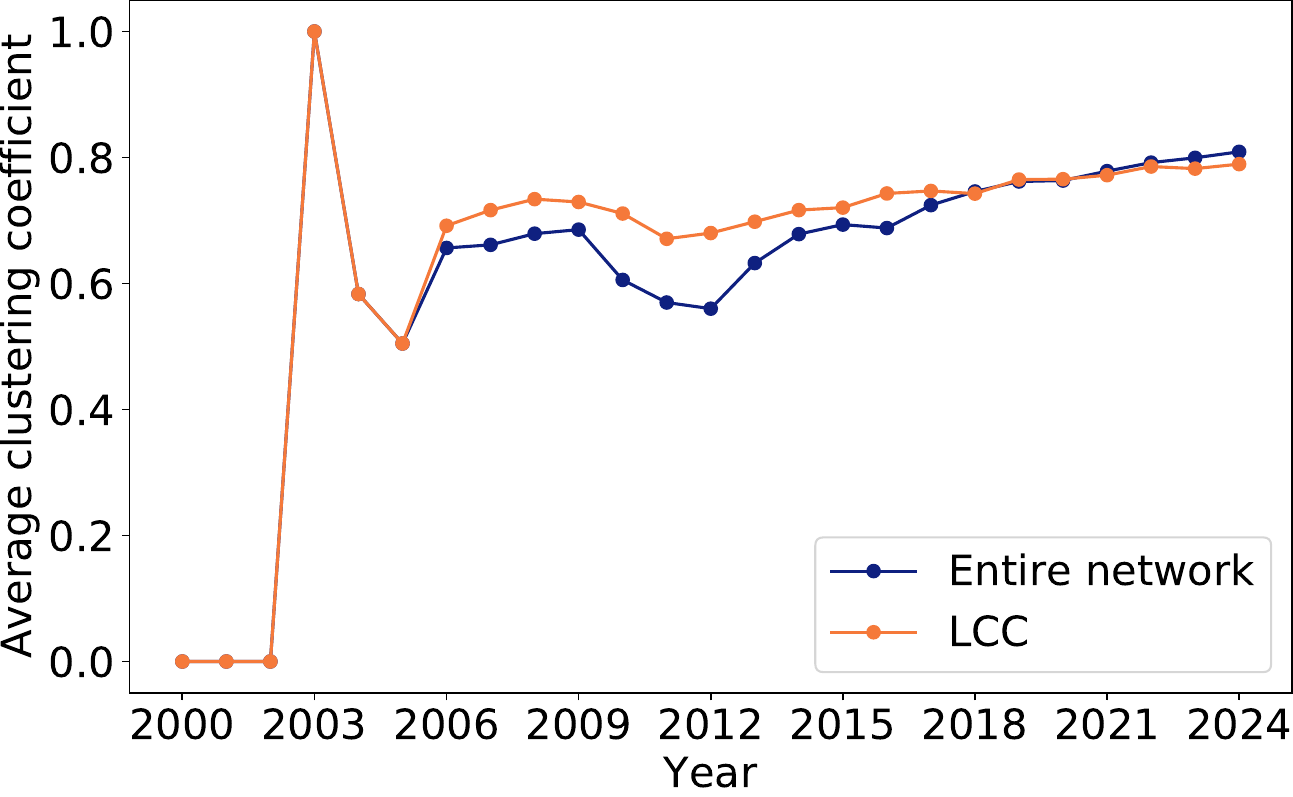}
    \caption{Clustering coefficient per year.}
    \label{fig:clustering}
\end{figure}

The clustering coefficient of the peridynamics co-authorship network is defined as the probability (computed as a proportion) that two co-authors of a given author also co-authored a peridynamics publication with each other. Figure \ref{fig:clustering} shows the evolution of the clustering coefficient of the entire network compared to that of the LCC. The clustering coefficient has been increasing for both the LCC and the entire network for most later years.

\begin{figure}[h!]
    \centering
    \begin{subfigure}[b]{0.45\textwidth}
    \centering
    \includegraphics[width=\linewidth]{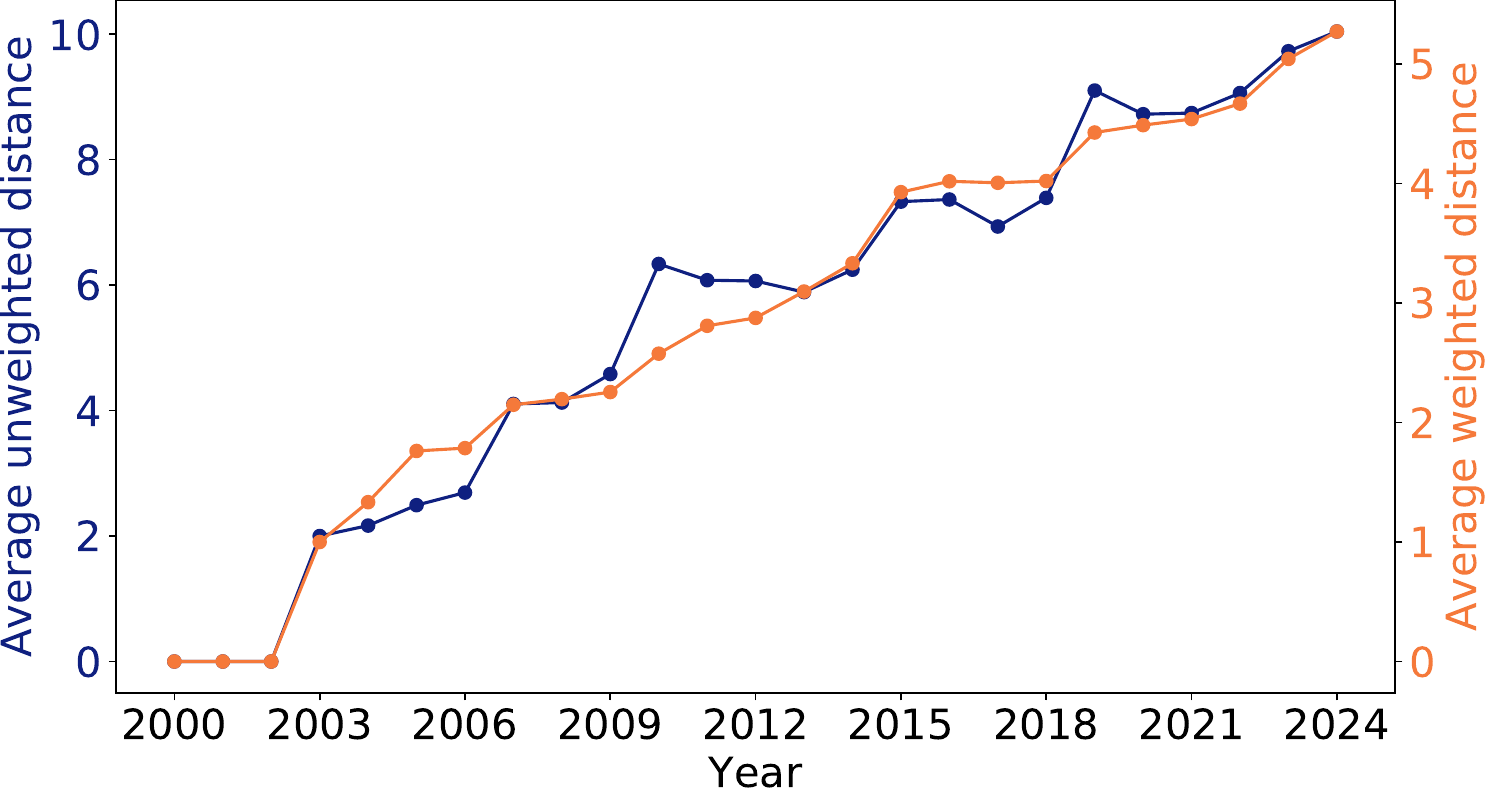}
    \caption{Average distance}
    \label{fig:distance and diameter a}
     \end{subfigure}
\;
  \begin{subfigure}[b]{0.465\textwidth}
    \centering
    \includegraphics[width=\linewidth]{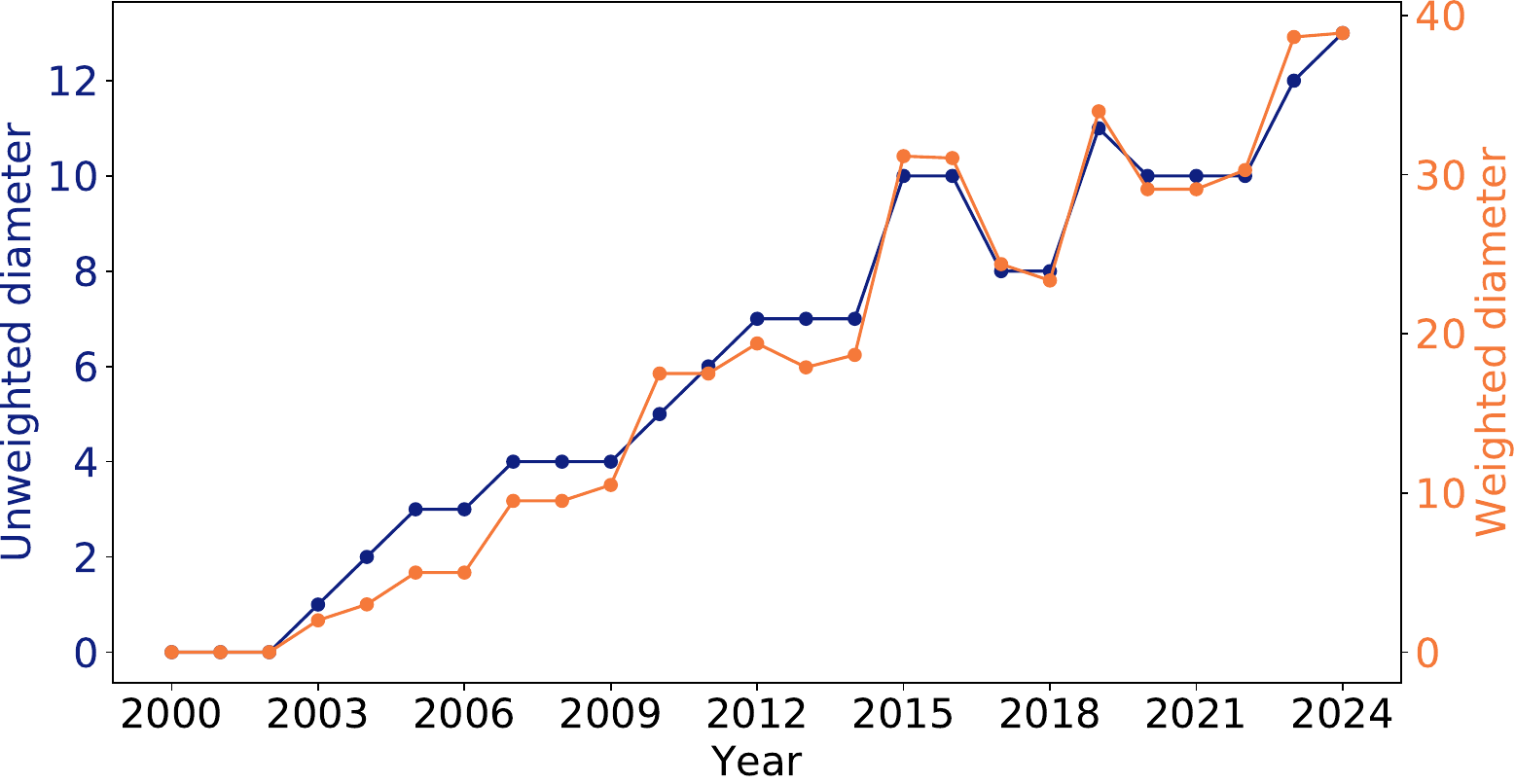}
    \caption{Diameter}
    \label{fig:distance and diameter b}
    \end{subfigure}
    \caption{Average distance and diameter of the LCC per year.}
    \label{fig:distance and diameter}
\end{figure}

The (unweighted) distance between two nodes in the network is defined as the length of the shortest path connecting the two nodes. One link is considered to have a distance (i.e., unweighted distance) of 1. The weighted distance is computed similarly, but the length of each link is considered to be the reciprocal of the link weight. Figure~\ref{fig:distance and diameter a} shows the evolution of the average distance and average weighted distance of the LCC over the years. The unweighted diameter (resp. weighted diameter) of a network is the longest length (resp. weighted length) among all shortest paths in the network. Figure~\ref{fig:distance and diameter b} shows the evolution of the unweighted and weighted diameter of the LCC over the years.

\begin{figure}[h!]
    \centering
    \includegraphics[width=0.5\linewidth]{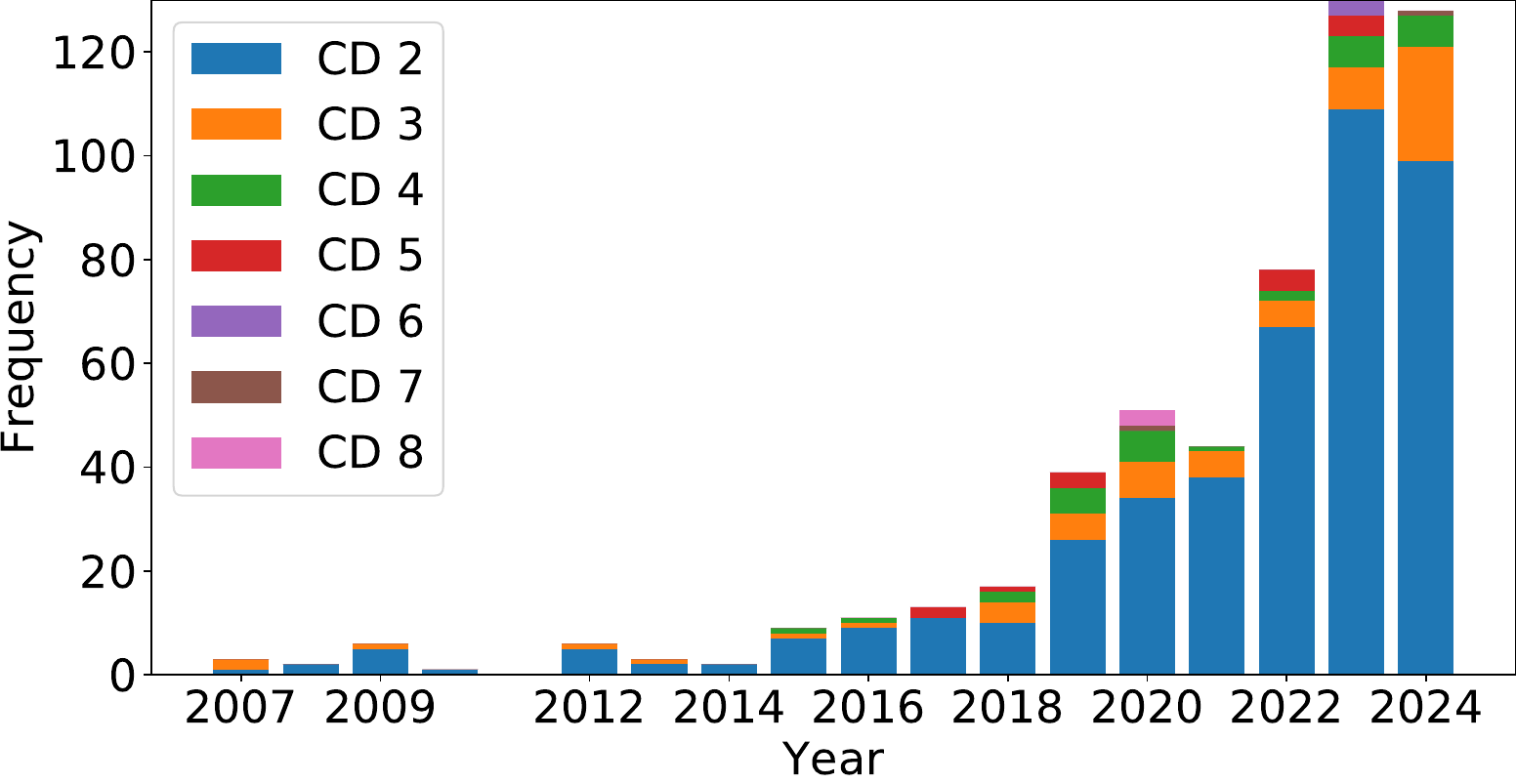}
    \caption{Intralinks by collaboration distance (CD) per year.}
    \label{fig:intralink}
\end{figure}

An intralink is defined as a link that connects two nodes in a connected component in a given year such that the two nodes were in the same connected component but not linked in the previous year. In our case, each intralink can be interpreted as a collaboration (co-authorship) between peridynamics scientists who were indirectly linked (i.e., there previously existed a path of pairwise co-authoring scientists connecting the two), when considering peridynamics publications. The unweighted distance of two scientists connected by an  intralink prior to the formation of the intralink is called the collaboration distance. Figure~\ref{fig:intralink} shows the distribution of collaboration distances over the years in a stacked bar plot, considering all intralinks formed between 2001 and 2024, grouped by collaboration distance (the plot starts at 2007 because no intralinks were formed before then). There were no intralinks in 2011.

\subsubsection{Other metrics discussion}
The clustering coefficient of the entire peridynamics co-authorship network increased from 0.69 in 2016 to 0.81 in 2024. The clustering coefficient of the LCC increased from 0.74 to 0.79 in the same period (see Figure~\ref{fig:clustering}). The high clustering value is not surprising given that the mean authorship team size for the peridynamics co-authorship network is 3.48 (see Section~\ref{sec:datasize}), which means that we often have trios or larger groups of co-authoring scientists joining the network as sets of fully-connected nodes, which would lead to a higher clustering coefficient.

The average weighted distance increases since 2019, but the average unweighted distance actually decreases from 2019 to 2021 before increasing again (see Figure~\ref{fig:distance and diameter a}). The weighted and unweighted diameters show a similar trend in recent years (see Figure~\ref{fig:distance and diameter b}); both diameters decreased from 2019 to 2020, stayed the same or only slightly increased from 2020 to 2022, significantly increased in 2023, and increased again in 2024.

The majority of intralinks have a collaboration distance of two (see Figure~\ref{fig:intralink}), which is not unexpected since many collaborations are likely facilitated by a mutual collaborator. On the other hand, we notice that there have been many intralinks with a collaboration distance of five or higher since 2017 (see Figure~\ref{fig:intralink}). Finally, we observe a slight decrease in the number of intralinks from 2023 to 2024 (130 to 128).

\section{Node-level Analysis}\label{sec:nodelevel}
In this section, we examine node-level metrics for the peridynamics co-authorship network, specifically using node centrality. For further discussion on node-level metrics, see \cite{dahal2023evolution}.

\subsection{Centrality}
The centrality of a node in a network is a measure of how ``connected'' the node is to all other nodes in the network. For co-authorship networks, centrality can be used to determine the most collaborative scientists in a field. Since there is no direct way to fairly compare centrality metrics between nodes in different connected components, we only compute and rank the centrality of the nodes in the LCC. We use four different centrality metrics:

\begin{enumerate}
    \item \textbf{Degree} centrality of a node is equal to the number of links of the node. It is equal to the number of co-authors of the corresponding scientist.
    \item \textbf{Weighted Degree} centrality of a node is equal to the sum of the weights on all the links of the node. It is equal to the number of co-authored publications of the corresponding  scientist.
    \item \textbf{Current Flow (C.F.) Closeness} centrality of a node quantifies how ``close'' the node is to all other nodes in the network, when considering all possible paths inversely weighted by length.
    \item \textbf{Current Flow (C.F.) Betweenness} centrality of a node quantifies the number of paths in the network that pass through the node, which indicates how ``in-between'' the node is to all other nodes in the network.
\end{enumerate}
For more details on the computation of these centrality metrics, see \cite{dahal2023evolution}. Table~\ref{table:centrality2024} shows the rankings of the peridynamics scientists corresponding to the four centrality metrics for 2024, displaying the top 20 ranked scientists. Note these rankings only consider peridynamics publication co-authorship; they do not reflect the significance of scientific contributions to the field. 
\begin{table*}[h!]
\centering
\begin{tabular}{rllll}
\toprule
 Rank &                  Degree &      Weighted Degree &          C.F. Closeness &             C.F. Betweenness \\
\midrule
    1 &             Oterkus, E. &          Madenci, E. &        Madenci, E. &             Madenci, E. \\
    2 &             Madenci, E. &          Oterkus, E. &        Oterkus, E. &             Silling, S. \\
    3 &             Silling, S. &    Oterkus, S. &        Silling, S. &             Oterkus, E. \\
    4 &              Bobaru, F. &           Bobaru, F. &  Oterkus, S. &              Bobaru, F. \\
    5 &                  Li, S. &     Huang, D.$^\dag$ &           Phan, N. &       Oterkus, S. \\
    6 &       Oterkus, S. &   Silling, S.$^\dag$ &          Barut, A. &                  Li, S. \\
    7 &         Galvanetto, U.$^\dag$ &             Zhou, X. &         Bobaru, F. &               Huang, D. \\
    8 &  Huang, D.$^\dag$ &            Guven, I. &      Diyaroglu, C. &              Askari, E. \\
    9 &  Zaccariotto, M.$^\dag$ &        Galvanetto, U.$^*$ &          Guven, I. &             Seleson, P. \\ 
   10 &              Foster, J. &  Zaccariotto, M.$^*$ &          Zhang, Q. &                 Han, F. \\
   11 &               Guven, I. & Zhang, Q.   $^*$ &       Dorduncu, M. &               Zhang, Q. \\
   12 &                 Han, F. &               Li, S. &         Askari, E. &              Foster, J. \\
   13 &               Zhang, Q. &               Gu, X. &             Gu, X. &                  Du, Q. \\
   14 &             Seleson, P. &  Barut, A. $^\ddag$ &        Seleson, P. &             Rabczuk, T. \\
   15 &             Rabczuk, T. &    Foster, J.$^\ddag$ &           Yang, Z. &   Galvanetto, U.$^\dag$ \\
   16 &           Chen, Z.$^*$ &     Phan, N.$^\ddag$ &        Weckner, O. &  Zaccariotto, M.$^\dag$ \\
   17 &            Cheng, Z.$^*$ &               Du, Q. &             Hu, Y. &               Guven, I. \\
   18 &          Gu, X.$^\ddag$ &             Chen, Z. &          Huang, D. &                  Gu, X. \\
   19 &         Liu, L.$^\ddag$ &             Chao, W. &        Rabczuk, T. &                 Ren, B. \\
   20 &                Zhou, X. &              Han, F. &    Galvanetto, U. &             Shojaei, A. \\
\bottomrule
\end{tabular}
\caption{2024 centrality rankings. Adjacent names in the same column with the same superscript are tied.}
\label{table:centrality2024}
\end{table*}

We define the \textbf{composite ranking} of a node to be its worst rank among all of the four centrality values. Table~\ref{table:compositecentrality} shows the composite ranking of the peridynamics scientists from 2020 to 2024, displaying the top 20 ranked scientists. For each year, we indicate rank changes (relative to the previous year) with up-pointing green triangles for increases and down-pointing red triangles for decreases along with the number of ranks increased or decreased.

\begin{table*}[h!]
\centering
\tiny
\begin{tabular}{rlllll}
\toprule
 Rank &                                              2020 Composite &                                              2021 Composite &                                            2022 Composite &                                             2023 Composite &                                                2024 Composite \\
\midrule
    1 &                                                Madenci, E.  &                                                Madenci, E.  &                                              Madenci, E.  &       Oterkus, E. {\color[HTML]{008800} $\blacktriangle$1} &          Madenci, E. {\color[HTML]{008800} $\blacktriangle$1} \\
    2 &        Oterkus, E. {\color[HTML]{008800} $\blacktriangle$1} &                                                Oterkus, E.  &                                              Oterkus, E.  &   Madenci, E. {\color[HTML]{880000} $\blacktriangledown$1} &      Oterkus, E. {\color[HTML]{880000} $\blacktriangledown$1} \\
    3 &    Silling, S. {\color[HTML]{880000} $\blacktriangledown$1} &                                                Silling, S.  &                                              Silling, S.  &                                               Silling, S.  &                                                  Silling, S.  \\
    4 &  Oterkus, S. {\color[HTML]{008800} $\blacktriangle$3} &                                          Oterkus, S.  &                                        Oterkus, S.  &                                         Oterkus, S.  &                                            Oterkus, S.  \\
    5 &     Bobaru, F. {\color[HTML]{880000} $\blacktriangledown$1} &                                                 Bobaru, F.  &                                               Bobaru, F.  &                                                Bobaru, F.  &                                                   Bobaru, F.  \\
    6 &                                                Weckner, O.  &         Askari, E. {\color[HTML]{008800} $\blacktriangle$2} &        Zhang, Q. {\color[HTML]{008800} $\blacktriangle$4} &                                                 Zhang, Q.  &                                                    Zhang, Q.  \\
    7 &          Guven, I. {\color[HTML]{008800} $\blacktriangle$2} &        Seleson, P. {\color[HTML]{008800} $\blacktriangle$5} &                                              Seleson, P.  &         Guven, I. {\color[HTML]{008800} $\blacktriangle$4} &                                                    Guven, I.  \\
    8 &     Askari, E. {\color[HTML]{880000} $\blacktriangledown$3} &    Weckner, O. {\color[HTML]{880000} $\blacktriangledown$2} &   Askari, E. {\color[HTML]{880000} $\blacktriangledown$2} &  Galvanetto, U. {\color[HTML]{008800} $\blacktriangle$16} &               Gu, X. {\color[HTML]{008800} $\blacktriangle$6} \\
    9 &         Foster, J. {\color[HTML]{008800} $\blacktriangle$7} &        Lehoucq, R. {\color[HTML]{008800} $\blacktriangle$5} &       Foster, J. {\color[HTML]{008800} $\blacktriangle$2} &  Zaccariotto, M.  {\color[HTML]{008800} $\blacktriangle$16} &            Huang, D. {\color[HTML]{008800} $\blacktriangle$2} \\
   10 &          Zhang, Q. {\color[HTML]{008800} $\blacktriangle$2} &                                                  Zhang, Q.  &  Weckner, O. {\color[HTML]{880000} $\blacktriangledown$2} &   Seleson, P. {\color[HTML]{880000} $\blacktriangledown$3} &                                                  Seleson, P.  \\
   11 &           Phan, N. {\color[HTML]{008800} $\blacktriangle$2} &     Foster, J. {\color[HTML]{880000} $\blacktriangledown$2} &        Guven, I. {\color[HTML]{008800} $\blacktriangle$1} &         Huang, D. {\color[HTML]{008800} $\blacktriangle$5} &          Rabczuk, T. {\color[HTML]{008800} $\blacktriangle$7} \\
   12 &    Seleson, P. {\color[HTML]{880000} $\blacktriangledown$2} &      Guven, I. {\color[HTML]{880000} $\blacktriangledown$5} &  Lehoucq, R. {\color[HTML]{880000} $\blacktriangledown$3} &    Askari, E. {\color[HTML]{880000} $\blacktriangledown$4} &  Galvanetto, U. {\color[HTML]{880000} $\blacktriangledown$4} \\
   13 &             Du, Q. {\color[HTML]{008800} $\blacktriangle$7} &          Parks, M. {\color[HTML]{008800} $\blacktriangle$2} &                                                Parks, M.  &   Weckner, O. {\color[HTML]{880000} $\blacktriangledown$3} &  Zaccariotto, M.  {\color[HTML]{880000} $\blacktriangledown$4} \\
   14 &    Lehoucq, R. {\color[HTML]{880000} $\blacktriangledown$6} &             Hu, Y. {\color[HTML]{008800} $\blacktriangle$5} &           Du, Q. {\color[HTML]{008800} $\blacktriangle$3} &            Gu, X. {\color[HTML]{008800} $\blacktriangle$1} &           Foster, J. {\color[HTML]{008800} $\blacktriangle$2} \\
   15 &      Parks, M. {\color[HTML]{880000} $\blacktriangledown$4} &       Phan, N. {\color[HTML]{880000} $\blacktriangledown$4} &           Gu, X. {\color[HTML]{008800} $\blacktriangle$1} &      Shojaei, A. {\color[HTML]{008800} $\blacktriangle$15} &               Hu, Y. {\color[HTML]{008800} $\blacktriangle$5} \\
   16 &             Gu, X. {\color[HTML]{008800} $\blacktriangle$1} &                                                     Gu, X.  &        Huang, D. {\color[HTML]{008800} $\blacktriangle$2} &    Foster, J. {\color[HTML]{880000} $\blacktriangledown$7} &       Askari, E. {\color[HTML]{880000} $\blacktriangledown$4} \\
   17 &      Diyaroglu, C. {\color[HTML]{008800} $\blacktriangle$4} &         Du, Q. {\color[HTML]{880000} $\blacktriangledown$4} &     Phan, N. {\color[HTML]{880000} $\blacktriangledown$2} &   Lehoucq, R. {\color[HTML]{880000} $\blacktriangledown$5} &      Weckner, O. {\color[HTML]{880000} $\blacktriangledown$4} \\
   18 &     Gunzburger, M. {\color[HTML]{008800} $\blacktriangle$1} &          Huang, D. {\color[HTML]{008800} $\blacktriangle$4} &       Hu, Y. {\color[HTML]{880000} $\blacktriangledown$4} &      Rabczuk, T. {\color[HTML]{008800} $\blacktriangle$28} &            Chen, Z. {\color[HTML]{008800} $\blacktriangle$11} \\
   19 &         Hu, Y. {\color[HTML]{880000} $\blacktriangledown$5} &  Diyaroglu, C. {\color[HTML]{880000} $\blacktriangledown$2} &                                            Diyaroglu, C.  &            Xu, J. {\color[HTML]{008800} $\blacktriangle$2} &              Xia, X. {\color[HTML]{008800} $\blacktriangle$5} \\
   20 &         Ha, Y. {\color[HTML]{880000} $\blacktriangledown$5} &             Xu, J. {\color[HTML]{008800} $\blacktriangle$1} &   Gunzburger, M. {\color[HTML]{008800} $\blacktriangle$2} &        Hu, Y. {\color[HTML]{880000} $\blacktriangledown$2} &      Shojaei, A. {\color[HTML]{880000} $\blacktriangledown$5} \\
\bottomrule
\end{tabular}
\caption{Composite centrality rankings for 2020-2024.}
\label{table:compositecentrality}
\end{table*}

Each centrality metric represents a distinct facet of network influence: degree and weighted degree highlight, respectively, the number and strength of direct connections; C.F. closeness reflects overall reachability within the network; and C.F. betweenness identifies authors serving as bridges. In contrast, the composite ranking integrates these diverse measures into a single metric, offering a comprehensive view of an author's overall network prominence.

Tables~\ref{table:centrality2024} and \ref{table:compositecentrality} reveal that some authors consistently occupy top ranks across the centrality measures, highlighting their roles as network hubs. For example, in the composite rankings, the top five positions (occupied by Madenci, E., Oterkus, E., Silling, S., Oterkus, S., and Bobaru, F.) remain largely stable, although a slight variation occurs between 2022 and 2024: the top two positions exchange places in 2023 and then revert in 2024. Moreover, when examining the individual centrality metrics from 2024, four of these five names consistently appear in the top four positions of each ranking, albeit with some variations. Besides the top five ranked authors in the composite rankings, other authors such as Zhang, Q., Guven, I., Gu, X., Seleson, P., Weckner, O., Foster, J., Hu, Y., and Askari, E. consistently appear among the top 20 in the composite rankings, and several appear in at least three of the top 20 individual centrality metrics from 2024. However, in contrast to the largely steady rankings of the top five authors, some authors experienced a steady decrease or sudden drop in rank in recent years, while others experienced a steady increase or sudden jump in rank over the same period, suggesting evolving patterns of collaboration.

\section{Country-based Analysis}
\label{sec:country}
In this section, we analyze the peridynamics co-authorship network by considering the distribution of the countries of the authors' institutional affiliations. We were inspired to do this analysis due to the potential impact of the COVID-19 pandemic on the evolution of the peridynamics co-authorship network during 2020-2024. Other works in the literature analyzed the changes in scientific publishing that occurred since 2020, particularly when considering the COVID-19 pandemic \citep{deryugina2021covid, gao2021potentially}. In particular, some works in the literature established differences between the scholarly output of scientists in different countries during the pandemic. For example, \cite{bohm2023impact} noticed that the number of scientists entering the field of astronomy per year decreased during COVID-19 in every country except in Japan, Taiwan, and China. This raises the possibility that the co-authorship trends within the peridynamics community might have evolved differently across countries.

To attain results comparable to the literature, in this section we analyze the peridynamics co-authorship network through a country-based lens, where we additionally consider the countries of the authors' institutional affiliations. This is different from the network analysis presented in the previous sections, which has not used any additional information about the nodes (authors) in the network beyond network metrics. While there are many country-based analysis options, we focus the analysis on an anomalous network behavior around 2020, which can be seen in Figure \ref{fig:size}. We perform a basic country-based analysis of the nodes entering the network to see if there are any differences between different countries in terms of their impact on the peridynamics co-authorship network. We leave a thorough country-based analysis of the peridynamics co-authorship network for future work.

\begin{figure}[h!]
    \centering
    \includegraphics[width=0.8\linewidth]{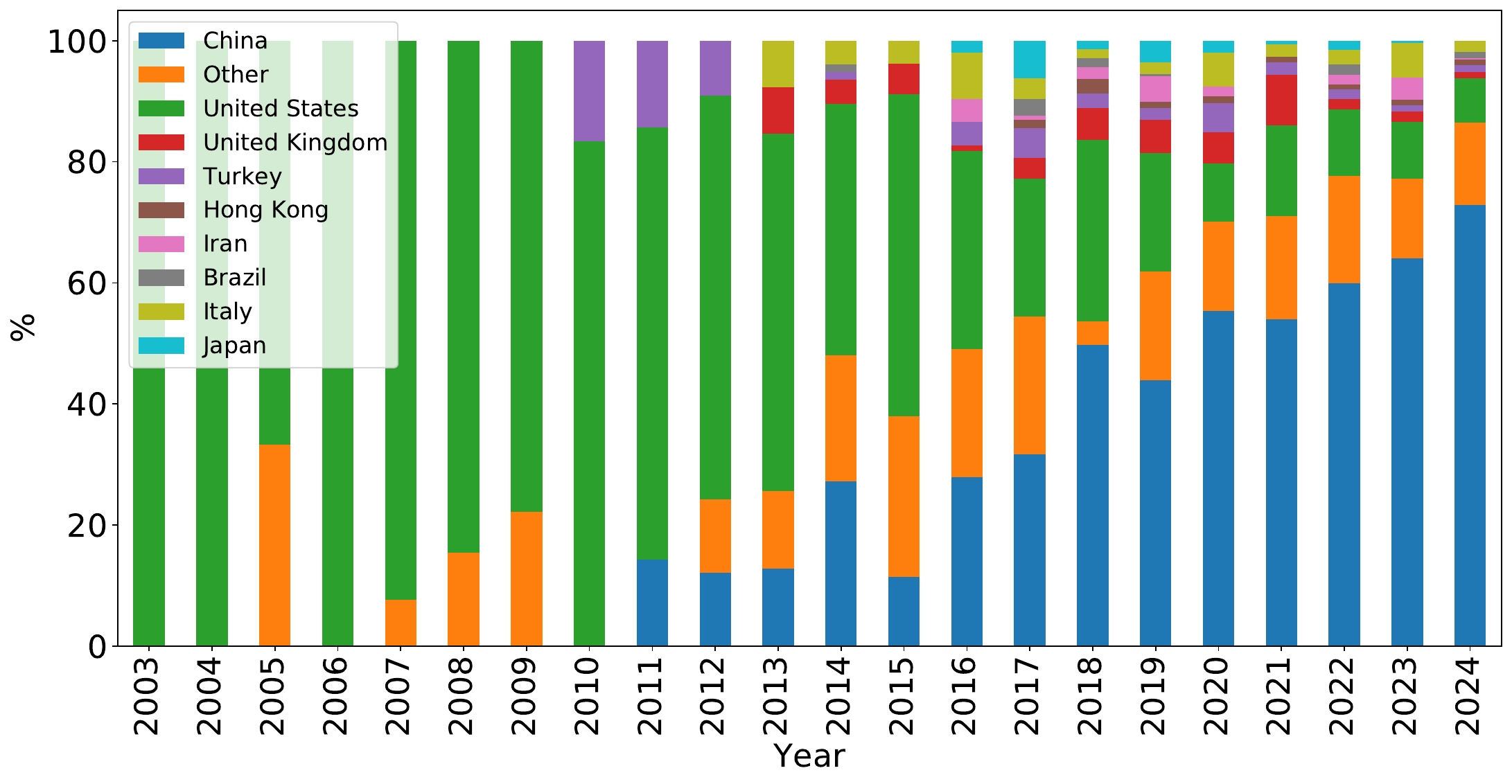}
    \caption{Country affiliation of new scientists joining the peridynamics co-authorship network per year. The countries in the legend are sorted by their values in descending order for the year 2024. ``Other'' refers to an institutional affiliation in any country that is not explicitly listed in the legend.}
    \label{fig:nodecountries}
\end{figure}

We begin by examining the countries of the authors' institutional affiliations of the new nodes joining the peridynamics co-authorship network each year. For each publication in the publication-author list, we assign a ``country affiliation'' to each author based on the country listed for them in the publication. 
For simplicity, we only consider the first affiliation listed for each author in the dataset. While the Scopus database is quite comprehensive, the affiliations column for a few records was incomplete in the Scopus data that we downloaded. Specifically, we had to manually inspect the original publication records and add the missing country affiliations for 17 publications. In the publication-author list, there was only one country affiliation for every author in the year that they joined, so there is no ambiguity in this assignment. Figure \ref{fig:nodecountries} is a stacked bar plot showing the distribution of country affiliations of new nodes for each year. Since 2020, the majority of scientists joining the network are affiliated with Chinese institutions.

\begin{figure}
    \centering    \includegraphics[width=0.5\linewidth]{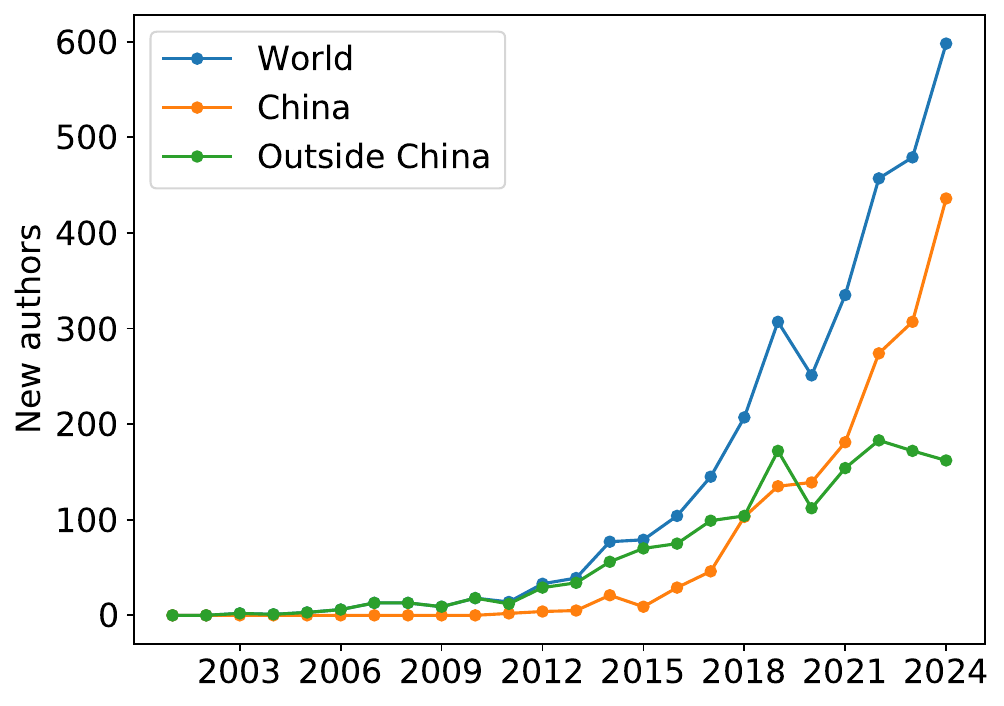}
    \caption{Number of new peridynamics authors per year, comparing those affiliated with Chinese institutions to those associated with institutions outside China.}
    \label{fig:comparechina}
\end{figure}

Because the majority of new nodes in the peridynamics co-authorship network since 2020 correspond to authors from Chinese institutions, we examine the distribution of new authors joining the network from Chinese institutions versus those joining from institutions outside China. Figure~\ref{fig:comparechina} shows that there is no decrease in the number of new authors joining the peridynamics co-authorship network when looking only at Chinese institutions, even after 2020. In addition, in every year since 2020, more nodes representing scientists from Chinese institutions join the network than nodes representing scientists associated with institutions from all other countries combined. 

We observed a qualitative change in the peridynamics co-authorship network since 2020: the majority of nodes joining the network represent scientists affiliated with Chinese institutions. When looking at peridynamics co-authorship trends for the entire world except for China, we see a dip in 2020, but no such dip is apparent when only considering new authors affiliated with Chinese institutions. This country-dependent difference in scholarly output is consistent with similar studies of scientific collaboration. For example, \cite{garcia2024silver} observed a difference in publication patterns in the early 2020s depending on the country (though the change in publication patterns also varied based on the research field).

\section{Summary}\label{sec:summary}
We examined the evolution of the peridynamics co-authorship network over the past quarter century, extending our previous study \citep{dahal2023evolution} with a  focus on 2020–2024 by employing three types of analysis: network-level, node-level, and country-based.

The network-level analysis quantified the peridynamics community’s rapid growth—publications have more than doubled and the number of authors has tripled since 2019 despite a slight recent slowdown. It shows that the community has a core group of scientists, represented by the largest connected component (LCC), which consistently comprises roughly half of the network in recent years and continues to grow relative to the other, significantly smaller connected components. Additionally, recent trends reveal a rise in the network's clustering coefficient and a more-than-twofold increase in large co-authorship teams (five or more authors) during 2020–2024 relative to the period up to 2019, even as the most common team size remains three. At the same time, there is an increasing trend in network distances and diameters, with more new collaborations occurring among scientists who are further apart in the network (with a collaboration distance of five or more), even though most new collaborations still occur at a distance of two. The node-level analysis identified the most collaborative peridynamics scientists by ranking them based on selected network metrics. It reveals that some authors consistently ranked on top positions while demonstrating dynamic patterns of collaboration overall with variations depending on the selected metric and fluctuations in composite rankings over the years. The country-based analysis indicated that since 2020, most new authors have been affiliated with Chinese institutions, with China alone contributing more new nodes to the network each year than all other countries combined. Furthermore, the analysis revealed that while worldwide growth in the peridynamics community dropped in 2020, growth remained steady when considering only Chinese institutions, consistent with literature observations on the COVID-19 pandemic's impact on scientific collaboration.

\section*{Acknowledgements}
B. Dahal was supported by the U.S. Department of Energy, Office of Science, Office of Workforce Development for Teachers and Scientists (WDTS) under the Science Undergraduate Laboratory Internship program. P. Seleson was supported by the Laboratory Directed Research and Development Program of Oak Ridge National Laboratory, managed by UT-Battelle, LLC, for the U.S. Department of Energy.

\bibliography{sn-bibliography}

\begin{thebibliography}{}
\renewcommand{\doi}[1]{\url{https://doi.org/#1}}
\bibcommenthead

\bibitem [\protect \citeauthoryear {%
B{\"o}hm%
\ \BBA {} Liu%
}{%
B{\"o}hm%
\ \BBA {} Liu%
}{%
{\protect \APACyear {2023}}%
}]{%
bohm2023impact}
\APACinsertmetastar {%
bohm2023impact}%
\begin{APACrefauthors}%
B{\"o}hm, V.%
\BCBT {}\ \BBA {} Liu, J.%
\end{APACrefauthors}%
\unskip\
\newblock
\APACrefYearMonthDay{2023}{}{}.
\newblock
{\BBOQ}\APACrefatitle {Impact of the {COVID-19} pandemic on publishing in
  astronomy in the initial two years} {Impact of the {COVID-19} pandemic on
  publishing in astronomy in the initial two years}.{\BBCQ}
\newblock
\APACjournalVolNumPages{Nature Astronomy}{7}{1}{105-112,}
\newblock
\begin{APACrefDOI} \doi{10.1038/s41550-022-01830-9} \end{APACrefDOI}
\newblock

\PrintBackRefs{\CurrentBib}

\bibitem [\protect \citeauthoryear {%
{Clarivate Analytics}%
}{%
{Clarivate Analytics}%
}{%
{\protect \APACyear {2024}}%
}]{%
wos}
\APACinsertmetastar {%
wos}%
\begin{APACrefauthors}%
{Clarivate Analytics}%
\end{APACrefauthors}%
\unskip\
\newblock
\APACrefYearMonthDay{2024}{}{}.
\newblock
\APACrefbtitle {{Web of Science}.} {{Web of Science}.}
\newblock
\APACrefnote{\url{https://www.webofknowledge.com/}}
\PrintBackRefs{\CurrentBib}

\bibitem [\protect \citeauthoryear {%
Dahal%
\ \BBA {} Seleson%
}{%
Dahal%
\ \BBA {} Seleson%
}{%
{\protect \APACyear {2025}}%
}]{%
dahal_2025_16897786}
\APACinsertmetastar {%
dahal_2025_16897786}%
\begin{APACrefauthors}%
Dahal, B.%
\BCBT {}\ \BBA {} Seleson, P.%
\end{APACrefauthors}%
\unskip\
\newblock
\APACrefYearMonthDay{2025}{}{}.
\newblock
\APACrefbtitle {{PDnetwork}.} {{PDnetwork}.}
\newblock
\APACaddressPublisher{}{Zenodo}.
\newblock
\APACrefnote{\url{https://doi.org/10.5281/zenodo.16897786}}
\PrintBackRefs{\CurrentBib}

\bibitem [\protect \citeauthoryear {%
Dahal%
, Seleson%
\BCBL {}\ \BBA {} Trageser%
}{%
Dahal%
\ \protect \BOthers {.}}{%
{\protect \APACyear {2023}}%
}]{%
dahal2023evolution}
\APACinsertmetastar {%
dahal2023evolution}%
\begin{APACrefauthors}%
Dahal, B.%
, Seleson, P.%
\BCBL {} Trageser, J.%
\end{APACrefauthors}%
\unskip\
\newblock
\APACrefYearMonthDay{2023}{}{}.
\newblock
{\BBOQ}\APACrefatitle {The evolution of the peridynamics co-authorship network}
  {The evolution of the peridynamics co-authorship network}.{\BBCQ}
\newblock
\APACjournalVolNumPages{Journal of Peridynamics and Nonlocal
  Modeling}{5}{3}{311--355,}
\newblock
\begin{APACrefDOI} \doi{10.1007/s42102-022-00082-5} \end{APACrefDOI}
\newblock

\PrintBackRefs{\CurrentBib}

\bibitem [\protect \citeauthoryear {%
Deryugina%
, Shurchkov%
\BCBL {}\ \BBA {} Stearns%
}{%
Deryugina%
\ \protect \BOthers {.}}{%
{\protect \APACyear {2021}}%
}]{%
deryugina2021covid}
\APACinsertmetastar {%
deryugina2021covid}%
\begin{APACrefauthors}%
Deryugina, T.%
, Shurchkov, O.%
\BCBL {} Stearns, J.%
\end{APACrefauthors}%
\unskip\
\newblock
\APACrefYearMonthDay{2021}{}{}.
\newblock
{\BBOQ}\APACrefatitle {{COVID-19} disruptions disproportionately affect female
  academics} {{COVID-19} disruptions disproportionately affect female
  academics}.{\BBCQ}
\newblock
 \APACrefbtitle {{AEA Papers and Proceedings}} {{AEA Papers and Proceedings}}\
  (\BVOL~111, \BPGS\ 164--168).
\PrintBackRefs{\CurrentBib}

\bibitem [\protect \citeauthoryear {%
{Elsevier}%
}{%
{Elsevier}%
}{%
{\protect \APACyear {2025}}%
}]{%
scopus}
\APACinsertmetastar {%
scopus}%
\begin{APACrefauthors}%
{Elsevier}%
\end{APACrefauthors}%
\unskip\
\newblock
\APACrefYearMonthDay{2025}{}{}.
\newblock
\APACrefbtitle {Scopus.} {Scopus.}
\newblock
\begin{APACrefURL} {https://www.scopus.com/} \end{APACrefURL}
\newblock
\APACrefnote{Accessed: August 13, 2025}
\PrintBackRefs{\CurrentBib}

\bibitem [\protect \citeauthoryear {%
Fruchterman%
\ \BBA {} Reingold%
}{%
Fruchterman%
\ \BBA {} Reingold%
}{%
{\protect \APACyear {1991}}%
}]{%
fruchterman1991graph}
\APACinsertmetastar {%
fruchterman1991graph}%
\begin{APACrefauthors}%
Fruchterman, T.M.J.%
\BCBT {}\ \BBA {} Reingold, E.M.%
\end{APACrefauthors}%
\unskip\
\newblock
\APACrefYearMonthDay{1991}{}{}.
\newblock
{\BBOQ}\APACrefatitle {Graph drawing by force-directed placement} {Graph
  drawing by force-directed placement}.{\BBCQ}
\newblock
\APACjournalVolNumPages{Software: Practice and Experience}{21}{11}{1129--1164,}
\newblock
\begin{APACrefDOI} \doi{10.1002/spe.4380211102} \end{APACrefDOI}
\newblock

\PrintBackRefs{\CurrentBib}

\bibitem [\protect \citeauthoryear {%
Gao%
, Yin%
, Myers%
, Lakhani%
\BCBL {}\ \BBA {} Wang%
}{%
Gao%
\ \protect \BOthers {.}}{%
{\protect \APACyear {2021}}%
}]{%
gao2021potentially}
\APACinsertmetastar {%
gao2021potentially}%
\begin{APACrefauthors}%
Gao, J.%
, Yin, Y.%
, Myers, K.R.%
, Lakhani, K.R.%
\BCBL {} Wang, D.%
\end{APACrefauthors}%
\unskip\
\newblock
\APACrefYearMonthDay{2021}{}{}.
\newblock
{\BBOQ}\APACrefatitle {Potentially long-lasting effects of the pandemic on
  scientists} {Potentially long-lasting effects of the pandemic on
  scientists}.{\BBCQ}
\newblock
\APACjournalVolNumPages{Nature Communications}{12}{1}{6188,}
\newblock
\begin{APACrefDOI} \doi{10.1038/s41467-021-26428-z} \end{APACrefDOI}
\newblock

\PrintBackRefs{\CurrentBib}

\bibitem [\protect \citeauthoryear {%
Garc{\'\i}a-Costa%
, Grimaldo%
, Bravo%
, Mehmani%
\BCBL {}\ \BBA {} Squazzoni%
}{%
Garc{\'\i}a-Costa%
\ \protect \BOthers {.}}{%
{\protect \APACyear {2024}}%
}]{%
garcia2024silver}
\APACinsertmetastar {%
garcia2024silver}%
\begin{APACrefauthors}%
Garc{\'\i}a-Costa, D.%
, Grimaldo, F.%
, Bravo, G.%
, Mehmani, B.%
\BCBL {} Squazzoni, F.%
\end{APACrefauthors}%
\unskip\
\newblock
\APACrefYearMonthDay{2024}{}{}.
\newblock
{\BBOQ}\APACrefatitle {The silver lining of {COVID-19} restrictions: research
  output of academics under lockdown} {The silver lining of {COVID-19}
  restrictions: research output of academics under lockdown}.{\BBCQ}
\newblock
\APACjournalVolNumPages{Scientometrics}{129}{3}{1771--1786,}
\newblock
\begin{APACrefDOI} \doi{10.1007/s11192-024-04929-0} \end{APACrefDOI}
\newblock

\PrintBackRefs{\CurrentBib}

\bibitem [\protect \citeauthoryear {%
Hagberg%
, Schult%
\BCBL {}\ \BBA {} Swart%
}{%
Hagberg%
\ \protect \BOthers {.}}{%
{\protect \APACyear {2008}}%
}]{%
hagberg2008exploring}
\APACinsertmetastar {%
hagberg2008exploring}%
\begin{APACrefauthors}%
Hagberg, A.A.%
, Schult, D.A.%
\BCBL {} Swart, P.J.%
\end{APACrefauthors}%
\unskip\
\newblock
\APACrefYearMonthDay{2008}{}{}.
\newblock
{\BBOQ}\APACrefatitle {Exploring Network Structure, Dynamics, and Function
  using {NetworkX}} {Exploring network structure, dynamics, and function using
  {NetworkX}}.{\BBCQ}
\newblock
 \APACrefbtitle {Proceedings of the 7$^{\rm {th}}$ {Python in Science
  Conference (SciPy 2008)}} {Proceedings of the 7$^{\rm {th}}$ {Python in
  Science Conference (SciPy 2008)}}\ (\BPGS\ 11--15).
\PrintBackRefs{\CurrentBib}

\bibitem [\protect \citeauthoryear {%
Perianes-Rodriguez%
, Waltman%
\BCBL {}\ \BBA {} Van~Eck%
}{%
Perianes-Rodriguez%
\ \protect \BOthers {.}}{%
{\protect \APACyear {2016}}%
}]{%
perianes2016constructing}
\APACinsertmetastar {%
perianes2016constructing}%
\begin{APACrefauthors}%
Perianes-Rodriguez, A.%
, Waltman, L.%
\BCBL {} Van~Eck, N.J.%
\end{APACrefauthors}%
\unskip\
\newblock
\APACrefYearMonthDay{2016}{}{}.
\newblock
{\BBOQ}\APACrefatitle {Constructing bibliometric networks: A comparison between
  full and fractional counting} {Constructing bibliometric networks: A
  comparison between full and fractional counting}.{\BBCQ}
\newblock
\APACjournalVolNumPages{Journal of Informetrics}{10}{4}{1178--1195,}
\newblock
\begin{APACrefDOI} \doi{10.1016/j.joi.2016.10.006} \end{APACrefDOI}
\newblock

\PrintBackRefs{\CurrentBib}

\bibitem [\protect \citeauthoryear {%
Silling%
}{%
Silling%
}{%
{\protect \APACyear {2000}}%
}]{%
silling2000reformulation}
\APACinsertmetastar {%
silling2000reformulation}%
\begin{APACrefauthors}%
Silling, S.A.%
\end{APACrefauthors}%
\unskip\
\newblock
\APACrefYearMonthDay{2000}{}{}.
\newblock
{\BBOQ}\APACrefatitle {Reformulation of elasticity theory for discontinuities
  and long-range forces} {Reformulation of elasticity theory for
  discontinuities and long-range forces}.{\BBCQ}
\newblock
\APACjournalVolNumPages{Journal of the Mechanics and Physics of
  Solids}{48}{1}{175--209,}
\newblock
\begin{APACrefDOI} \doi{10.1016/S0022-5096(99)00029-0} \end{APACrefDOI}
\newblock

\PrintBackRefs{\CurrentBib}

\end{thebibliography}

\end{document}